\documentclass[12pt]{article}

\usepackage{scicite}

\usepackage{times}

\usepackage{bm}
\usepackage{amsfonts}
\usepackage{graphicx}
\usepackage{amsmath}
\usepackage{epsfig,graphics, color,bm}
\usepackage{cases}
\usepackage{subeqnarray}
\usepackage{mathrsfs,bm}
\usepackage[toc,page]{appendix}
\usepackage{graphicx,epsfig,amsmath,amsfonts,amscd}
\usepackage{float}
\usepackage{color}
\usepackage{latexsym,amssymb}
\usepackage{dsfont}
\usepackage{fancyhdr}
\usepackage{stmaryrd}
\usepackage{subfig}
\usepackage{float}
\usepackage{epstopdf}
\usepackage{comment}

\newcommand{\ignore}[1]{} %

\makeatletter
\newcommand\figcaption{\def\@captype{figure}\caption}
\makeatother

\newcommand{\beq}{\begin{equation}}
\newcommand{\eeq}{\end{equation}}
\newcommand{\beqn}{\begin{equation*}}
\newcommand{\eeqn}{\end{equation*}}
\newcommand{\be}{\begin{eqnarray}}
\newcommand{\ee}{\end{eqnarray}}
\newcommand{\bex}{\begin{eqnarray*}}
\newcommand{\eex}{\end{eqnarray*}}
\newcommand{\ba}{\begin{array}}
\newcommand{\ea}{\end{array}}

\newcommand{\0}{\boldsymbol{0}}
\definecolor{Red}{rgb}{1,0,0}
\definecolor{Blue}{rgb}{0,0,1}
\definecolor{Green}{rgb}{0,1,0}
\definecolor{magenta}{rgb}{1,0,0.6}
\definecolor{lightblue}{rgb}{0,0.5,1}
\definecolor{lightpurple}{rgb}{0.6,0.4,1}
\definecolor{gold}{rgb}{0.6,0.5,0}
\definecolor{orange}{rgb}{1,0.4,0}
\definecolor{hotpink}{rgb}{1,0,0.5}
\definecolor{newcolor2}{rgb}{0.5,0.3,0.5}
\definecolor{newcolor}{rgb}{0,0.3,1}
\definecolor{newcolor3}{rgb}{1,0,0.35}
\definecolor{darkgreen1}{rgb}{0,0.35, 0}
\definecolor{darkgreen}{rgb}{0,0.6, 0}
\definecolor{darkred}{rgb}{0.75,0,0}
\def\dps{\displaystyle}

\font\tenbi=cmmib10   at 11 pt
\font\sevenbi=cmmib10 at 9pt
\font\fivebi=cmmib7 at 6pt
\newfam\bifam
\textfont\bifam=\tenbi \scriptfont\bifam=\sevenbi
\scriptscriptfont\bifam=\fivebi

\font\tendb=msbm10 at 12 pt
\font\sevendb=msbm7
\newfam\dbfam
\textfont\dbfam=\tendb \scriptfont\dbfam=\sevendb

\def\0{{\bm0}}

\newenvironment{sciabstract}{%
\begin{quote} \bf}
{\end{quote}}

\title{A Universal Fractional Model of Wall-Turbulence}

\author
{Fangying Song$^{1},$ George Em Karniadakis$^{1\ast}$ \\
\\
\normalsize{$^{1}$Division of Applied Mathematics, Brown University,}
\normalsize{Providence, Rhode Island 02912, USA}\\
\\
\normalsize{$^\ast$Corresponding author; E-mail: george\_karniadakis@brown.edu}
}

\date{}

\begin{document} 
\graphicspath{{figures/}}


\maketitle


\begin{sciabstract}
Modeling of wall-bounded turbulent flows is still an open problem in classical physics, with only modest progress made in the last few decades beyond the so-called `log law', which describes only the intermediate region in wall-bounded turbulence, i.e., 
$30-50 y^+ \text{ to } 0.1-0.2 R^+$ (in wall units) in a pipe of radius $R$. Here we propose a fundamentally new approach based on fractional calculus to model the {\em entire} mean velocity profile from the wall to the centerline of the pipe. Specifically, we represent the Reynolds stresses with a non-local fractional derivative of {\em variable order} that decays with the distance from the wall. Surprisingly, we find that this variable fractional order has a universal form for all Reynolds numbers and for three different flow types, i.e., channel flow, Couette flow, and pipe flow. We first use existing data bases from direct numerical simulations (DNS) to learn the variable fractional order function, and subsequently we test it against other DNS data and experimental measurements, including the Princeton superpipe experiments. Taken together, our findings reveal the continuous and decaying change of rate of turbulent diffusion from the wall as well as the strong non-locality of turbulent interactions that intensify away from the wall.
\end{sciabstract}


Osborne Reynolds \cite{reynolds1895dynamical} was the first to describe turbulence statistically by decomposing the instantaneous velocity vector into an average field and its fluctuation. Upon substitution into the Navier-Stokes equations and averaging, assuming quasi-stationarity, a new modified equation emerges for the average velocity that includes an additional term, namely the averaged dissipation tensor leading to the so-called turbulence-closure problem \cite{pope2000turbulent}. Addressing the closure complexity has been a century long pursuit, starting with the seminal work of Ludwig Prandtl \cite{prandtl1925bericht}, who proposed a simplified mixing length model in analogy with Fick's law of local diffusion. Interestingly, at about the same time Richardson \cite{richardson1926atmospheric} in his attempt to unify turbulent diffusion with molecular diffusion he combined geophysical measurements with Brownian motion to produce his famous scaling law on turbulent pair diffusivity! While ingenious, both approaches assume implicitly {\em locality} in turbulent interactions, which limits the universality of the derived correlations -- an open standing question for over a century! As Kraichnan \cite{kraichnan1964direct} pointed out, Prandtl's approach is valid only when the spatial scale of inhomogeneity of the mean field is large compared to the mixing length. This assumption is clearly violated in most turbulent flows, e.g. in Reynolds' pipe flow, where the turbulent eddies are of the size of the pipe radius! This has motivated research in non-local constitutive equations of turbulence, and indeed Prandtl himself in subsequent work \cite{prandtl1942bemerkungen} developed a turbulent shear-layer model in an attempt to introduce non-locality in his approach. Kraichnan \cite{kraichnan1964direct}  pioneered such non-local approximations, and recently extensions of  the second Prandtl non-local model were   proposed in the literature \cite{egolf2016nonlocal}. Similarly, if Richardson had not used the Brownian motion data, his best fit would have given an exponent for the mean-square-displacement greater than 1.5 (indicative of non-locality) instead of the familiar exponent of $4/3$, which is consistent with local interactions.

Fractional calculus is an effective tool to solve complex problems with non-locality and scale-free self-similar processes as well as non-Gaussian statistics. L\'{e}vy statistics lead to anomalous diffusion \cite{carpinteri2014fractals} and can also model effectively turbulent intermittency \cite{shlesinger1987levy}. Hence, it is possible that the turbulent eddy diffusivity could be modeled accurately by fractional Reynolds stresses \cite{egolf2017fractional}. Based on physical arguments in order to represent non-locality and intermittency, Chen \cite{CW2006FT} proposed a fractional Laplacian as a model for representing the Reynolds stresses with a fixed fractional exponent $\alpha=2/3$. More recently, starting with the Boltzmann equation, Epps and Cushman-Roisin  \cite{Epps2017turbulence} derived rigorously the {\em fractional} Navier-Stokes equations by replacing the Maxwell-Boltzmann distribution with the more general L\'{e}vy $\alpha$-stable distribution. For $\alpha=2$ the new equations revert to the standard Navier-Stokes equations while for $\alpha=1$ one obtains the logarithmic velocity profile known as the law of the wall \cite{GWK2007}. The work in \cite{Epps2017turbulence} presents a new formulation for turbulence modeling that may lead to new fundamental understanding of turbulence but it is only valid in an open domain and thus ignores the important issue of non-local boundary conditions encountered in defining fractional Laplacians in bounded domains \cite{lischke2018fractional}.

For wall-bounded turbulence the effective {\em rate of diffusion} varies with the distance from the wall. Hence, we exploit the power of fractional calculus that allows {\em variable fractional order}, and we propose a variable-order fractional differential equation for modeling the Reynolds stresses, i.e., $\alpha(y)$, where $y$ is the distance from the wall.  In particular, here we consider fully-developed turbulent flows with {\em one-dimensional} (dimensionless) averaged velocity $U(y)=u/V$ ($V$ is the characteristic velocity), 
including channel flows, pipe flows and also Couette flows for which we will apply a unified 
fractional modeling approach. Specifically, assuming that the flow direction is along $x$ and $y$ is the wall-normal direction (distance from the wall), 
we consider the variable fractional model (VFM) in Cartesian coordinates (in non-dimensional form): 
\beq\label{FM}
\nu_0\frac{\partial^2U}{\partial y^2}+\nu(y)D_y^{\alpha(y)}U=f,\ \forall y\in\Lambda=(0,1], 
\eeq
with $\alpha(0)=1$, $0\leq\alpha(y)\leq1$, $D_y^\alpha$ is the (Caputo) fractional derivative, $f=\partial P/\partial x=1$ is the dimensionless pressure gradient,  $U(y)$ is the mean velocity we want to model, and $\nu_0$ is the kinematic viscosity (dimensionless). The Caputo derivative is defined as
\beqn
{}D_y^\alpha U(y)=\frac{1}{\Gamma(1-\alpha)}\int_0^y(y-\tau)^{-\alpha}U'(\tau)d\tau,
\eeqn
and it is identical to the Riemann-Liouville left-sided derivative because $U(0)=0$. 
Interestingly, we can obtain the scalar coefficient $\nu(y)$ (we will refer to it as turbulent diffusivity although it does not have the right units) explicitly in terms of the fractional order $\alpha(y)$ from:
\beq\label{visc}
\nu(y)=\lim\limits_{y_0\rightarrow\frac{1}{Re_\tau}}\frac{f}{D_s^{\alpha(y)}U\big|_{s=y_0}}=f\Gamma(2-\alpha(y))Re_\tau^{-\alpha(y)}V/u_\tau,
\eeq
where $Re_\tau=u_\tau R/\nu_0$ is the friction Reynolds number, $R$ is the radius of the pipe (or the half channel width),  and $u_\tau$ is the wall friction velocity, $u_\tau=\sqrt{\tau_w/\rho}$, where $\tau_w=\mu\partial U/\partial y|_{y=0}$ is the wall shear stress with $\mu$ the dynamic viscosity.

The VFM of equation \eqref{FM} with the corresponding coefficient from equation \eqref{visc} is a new fractional Reynolds averaged equation in steady-state; it can also be extended to three-dimensional steady-state turbulent flows. Unlike integer-calculus based models, where the eddy diffusivity is placed inside the divergence operator, i.e.
$\nabla \cdot [\nu(y) (\nabla U + \nabla^T U)]$, for momentum conservation, here the operator depends on the distance from the wall and such arguments are not valid. In \cite{Supplementary} we consider this model and investigate its properties but we show that it does not lead to universality.  
An alternative model with $1 \leq \alpha(y) \leq 2$ is given below and further discussed in \cite{Supplementary}, which leads to universality. 
In \cite{Supplementary} we present the method on how to solve the above equation for $\alpha(y), y\in[0,1]$
given the velocity profile $U(y)$ by setting up an optimization problem. We will assume for now that $U(y)$ is given by available data from experiments or direct numerical simulations (DNS) for the specific geometry and Reynolds number, $Re_\tau$, we consider.

\begin{figure}[H]
\centering
\subfloat[]{
\includegraphics[height=2.2in]{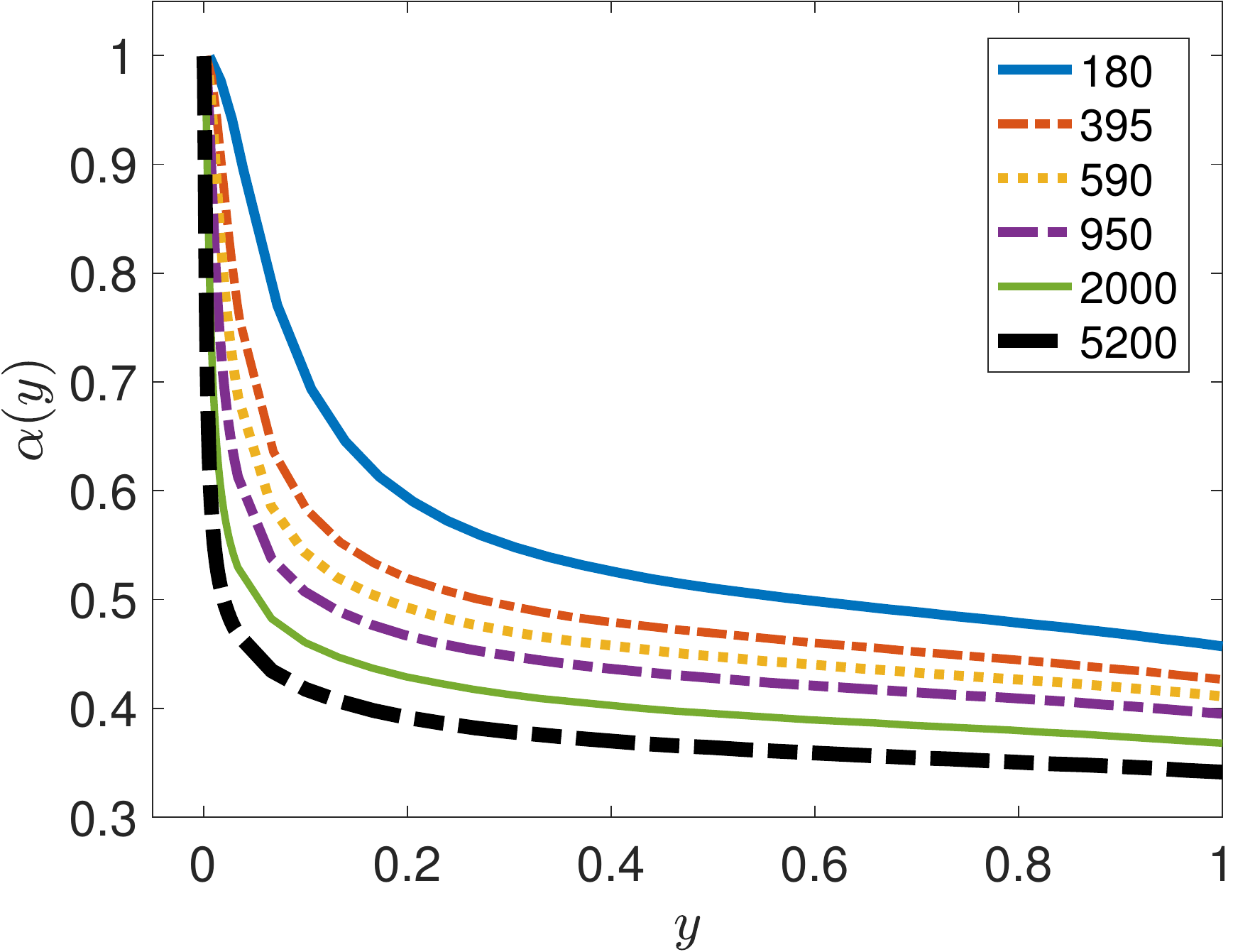}}
\subfloat[]{
\includegraphics[height=2.2in]{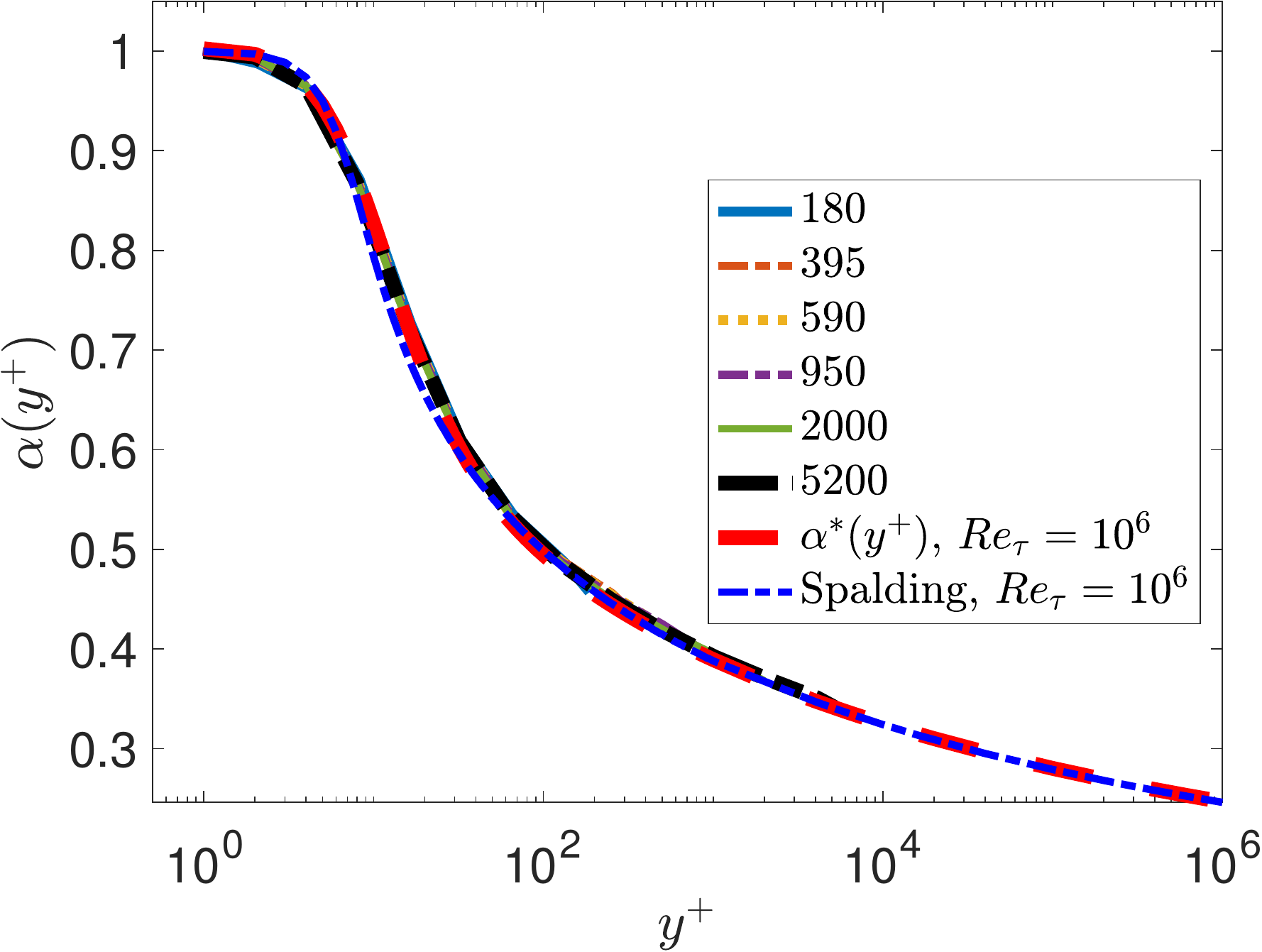}}\\
\caption{{\bf Learning the fractional variable-order $\alpha(y)$ using DNS data bases at $Re_\tau=180\ \text{to}\ 5200$}: (a) profiles of the fractional order $\alpha(y)$; (b) rescaled fractional order $\alpha(y^+)$ in viscous units.}
\label{alpha_y}
\end{figure}

\begin{figure}[H]
\centering
\subfloat[]{
\includegraphics[height=2.2in]{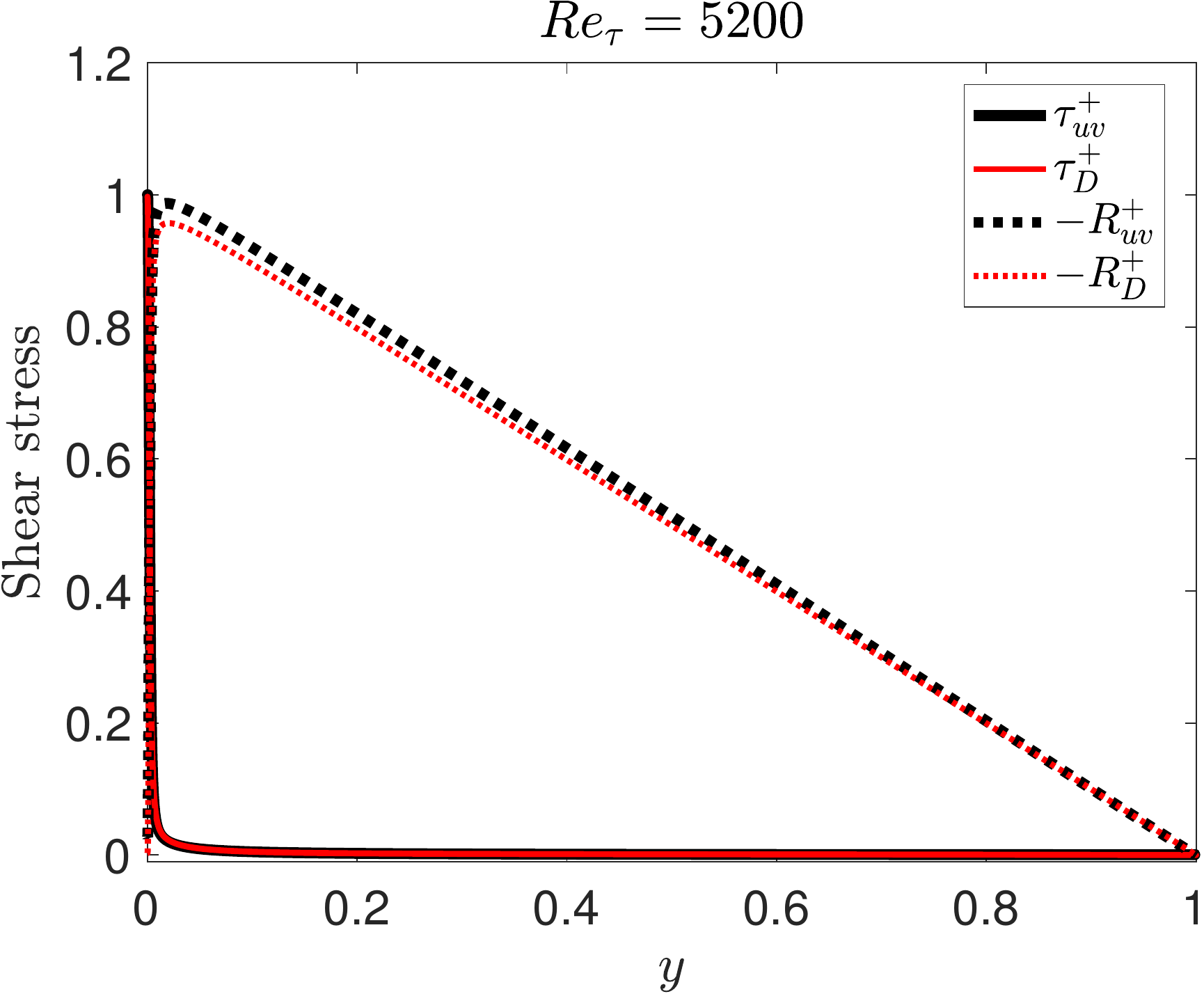}}
\subfloat[]{
\includegraphics[height=2.2in]{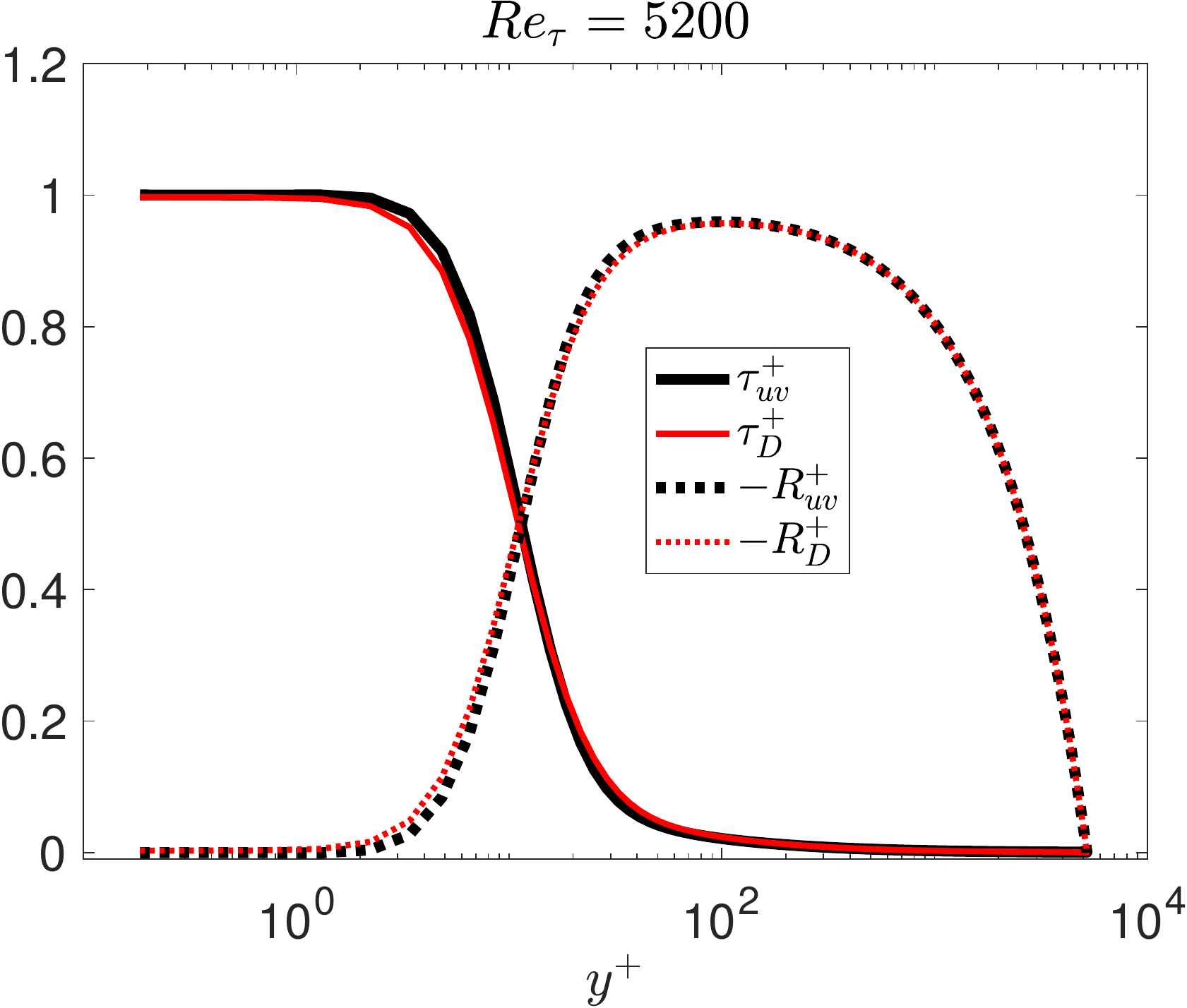}}
\caption{{\bf Accurate prediction of the shear stress at $Re_\tau=5200$ in outer units and wall units:} (a) Outer scaling; (b) Wall Units scaling.
Here $\tau_{uv}$ denotes the wall shear stress predicted by VFM (black-solid line) and  $\tau_D$ is the  corresponding profile from DNS data. $-R_{uv}$ (black-dash line) denotes the Reynolds shear stress predicted by VFM, and $-R_{D}$ is the corresponding profile from DNS data.}
\label{profile_stress5200}
\end{figure}

We first address turbulent {\em channel flow} for which DNS data are available up to $Re_\tau=5200$ \cite{LM2015D}. 
Solving for $\alpha(y)$, which uniquely determines the Reynolds stresses, we plot in Fig. \ref{alpha_y}(a) profiles of the fractional 
order $\alpha(y)$ for different $Re_\tau$ as a function of the non-dimensional distance from the wall $y\in[0,1]$. We see a strong dependence of $\alpha(y)$ on
$Re_\tau$, however, if we re-plot all data in terms of the viscous wall units, i.e. $y^+=y u_\tau/\nu_0$
we see a collapse of all results into a single universal curve, see Fig. \ref{alpha_y}(b). Moreover, we have employed the empirical Spalding formula \cite{white1991viscous} 
for $U^+ =u/u_\tau$ in order to extend the results up to high $Re_\tau=10^6$, and again we obtain a similar universal scaling with the exception of low $Re_\tau$ for which the Spalding formula is known to be somewhat inaccurate. Using the same data for $U(y)$ we show that the alternative model with $1 \leq \alpha(y) \leq 2$ also leads to the same type of universality, see 
\cite{Supplementary}. The specific fitting function $\alpha^*(y^+)$ for $\alpha(y^+)$ is given also in
\cite{Supplementary}.
 This is a remarkable result as it goes beyond the logarithmic profile and connects the viscous sub-layer with the buffer zone, the logarithmic profile and the wake region seamlessly. While at first it looks like a perfect fitting exercise, it has important consequences due to the non-local interpretation of the fractional derivative involved, i.e., it shows
that non-locality is stronger away from the wall and at high Reynolds numbers.
%
To evaluate the predictability of the universal scaling, we now solve the forward equation \eqref{FM} to obtain $U(y)$ at $Re_\tau=[4200,6000,8600]$, which are
cases not used in the training of the model for $\alpha(y^+)$. The results presented in \cite{Supplementary}
 are in good agreement with DNS and experimental data.
Moreover, we wanted to investigate how accurately we predict the Reynolds stresses since previously published models, e.g. \cite{chen1998camassa},
showed poor agreement with DNS. To this end, in Fig. \ref{profile_stress5200} we plot for $Re_\tau=5200$ the profiles in the shear stress budget 
as explained in \cite{Supplementary} 
and we see very good agreement between the DNS data and the new universal VFM. 

\begin{figure}[H]
\centering
\subfloat[]{
\includegraphics[height=2.2in]{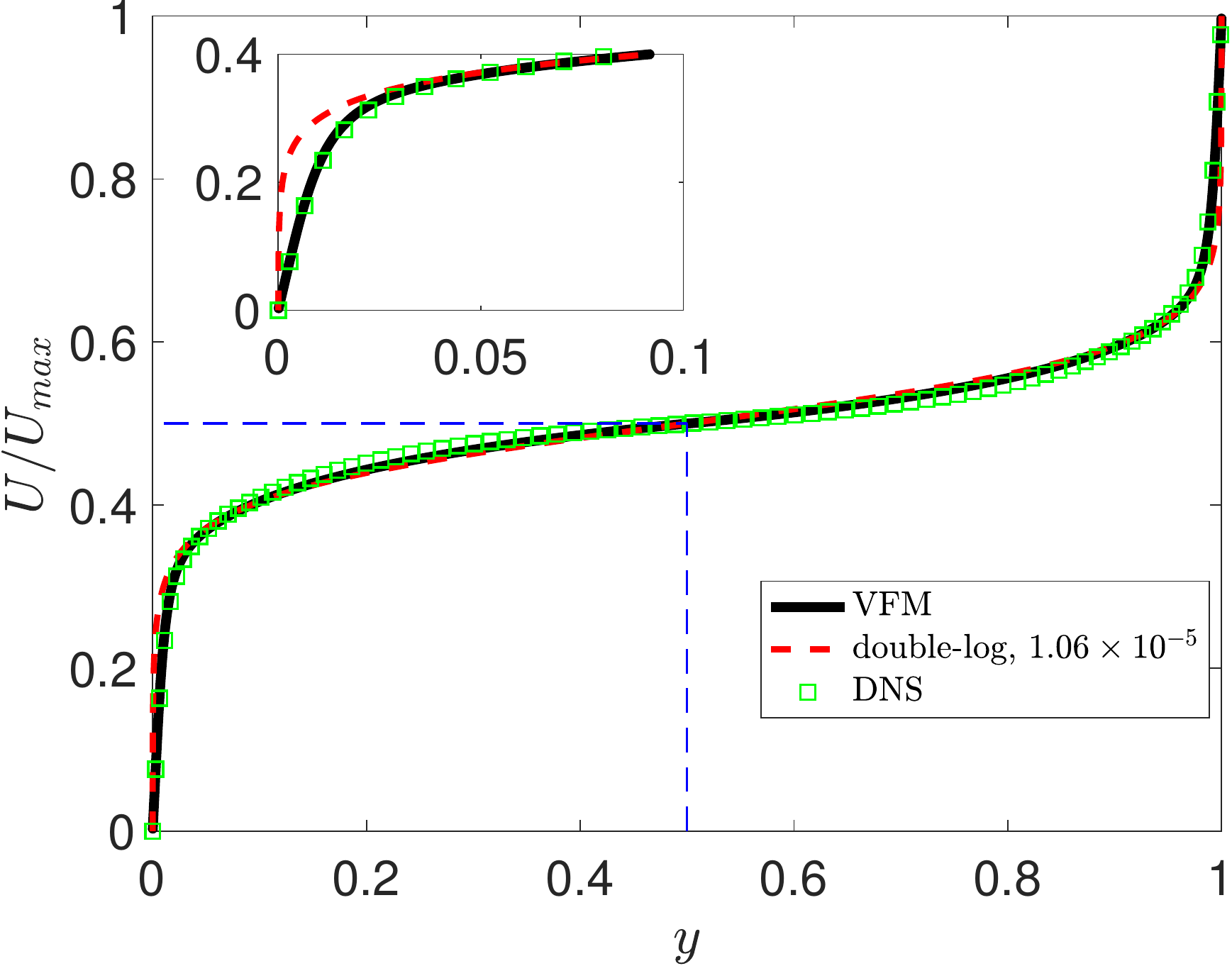}}
\subfloat[]{
\includegraphics[height=2.2in]{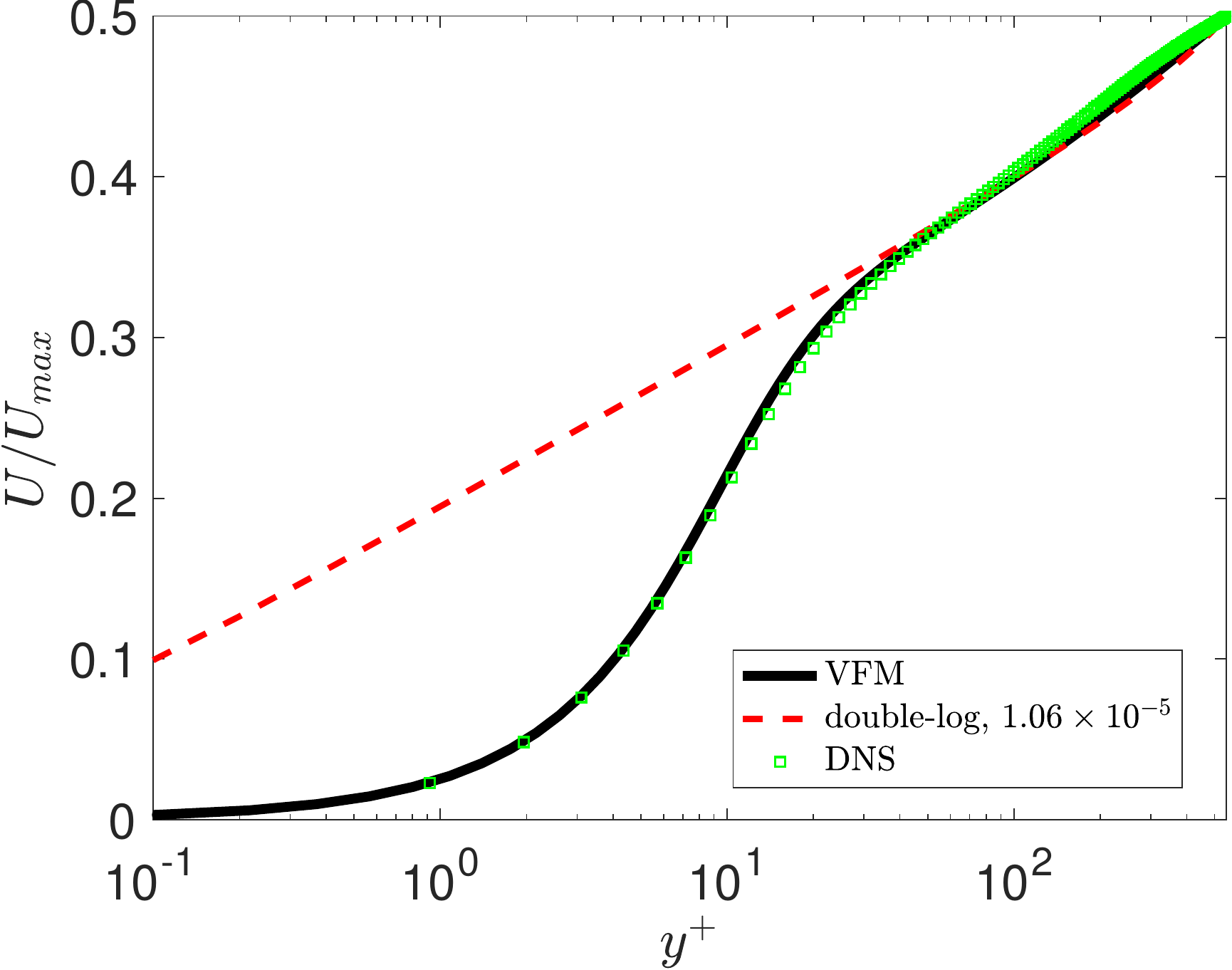}}
\figcaption{{\bf Predicting the Couette mean flow profile at $Re_\tau=550$}: (a) scaling with half channel width; (b) scaling in  wall units; VFM profiles are denoted by the black line, DNS data \cite{avsarkisov2014turbulent} by $\square$ symbols, and double-log profile by the red line.  The inset is
zoom in at the wall.}
\label{profile_y550}
\end{figure}

An even stricter test of predictability enabled with the universal VFM is to use it to predict $U(y)$ for a different type of flow, so next we consider turbulent {\em Couette flow}. A systematic study is presented in \cite{Supplementary}
but here we discuss a representative case for $Re_\tau=550$, see Fig. \ref{profile_y550}. We also include a 
double-log profile in the plot proposed in  \cite{Epps2017turbulence}, which is unable to capture the correct mean velocity unlike the prediction by the universal VFM; the DNS data are due to  \cite{avsarkisov2014turbulent}. The double-log layer profile has a free parameter and can be adjusted to improve the fitting, at least for certain Reynolds numbers as we show in \cite{Supplementary} 
but it cannot obtain good accuracy close to the wall.

\begin{figure}[http]
\centering
\subfloat[]{
\includegraphics[height=2.2in]{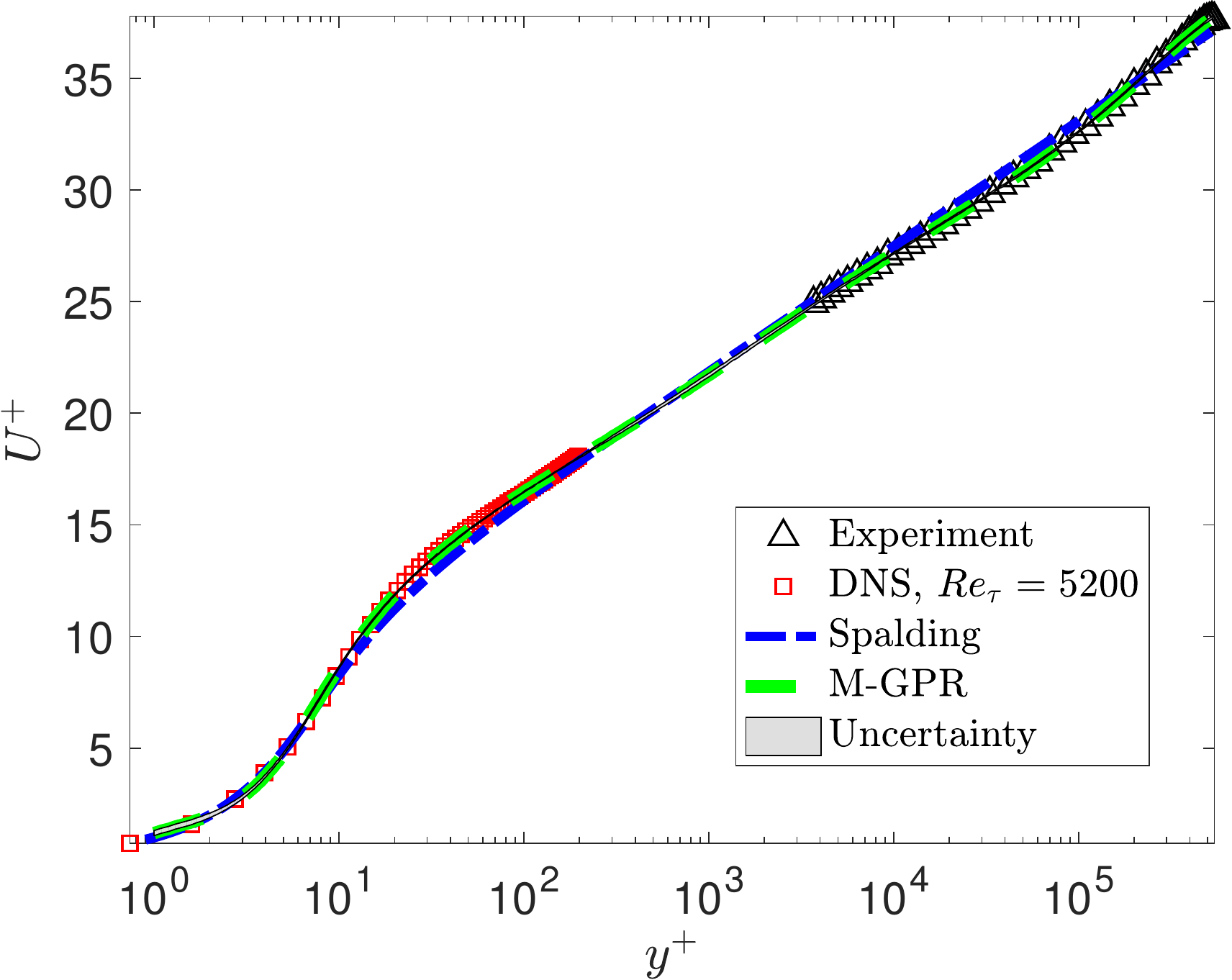}}
\subfloat[]{
\includegraphics[height=2.2in]{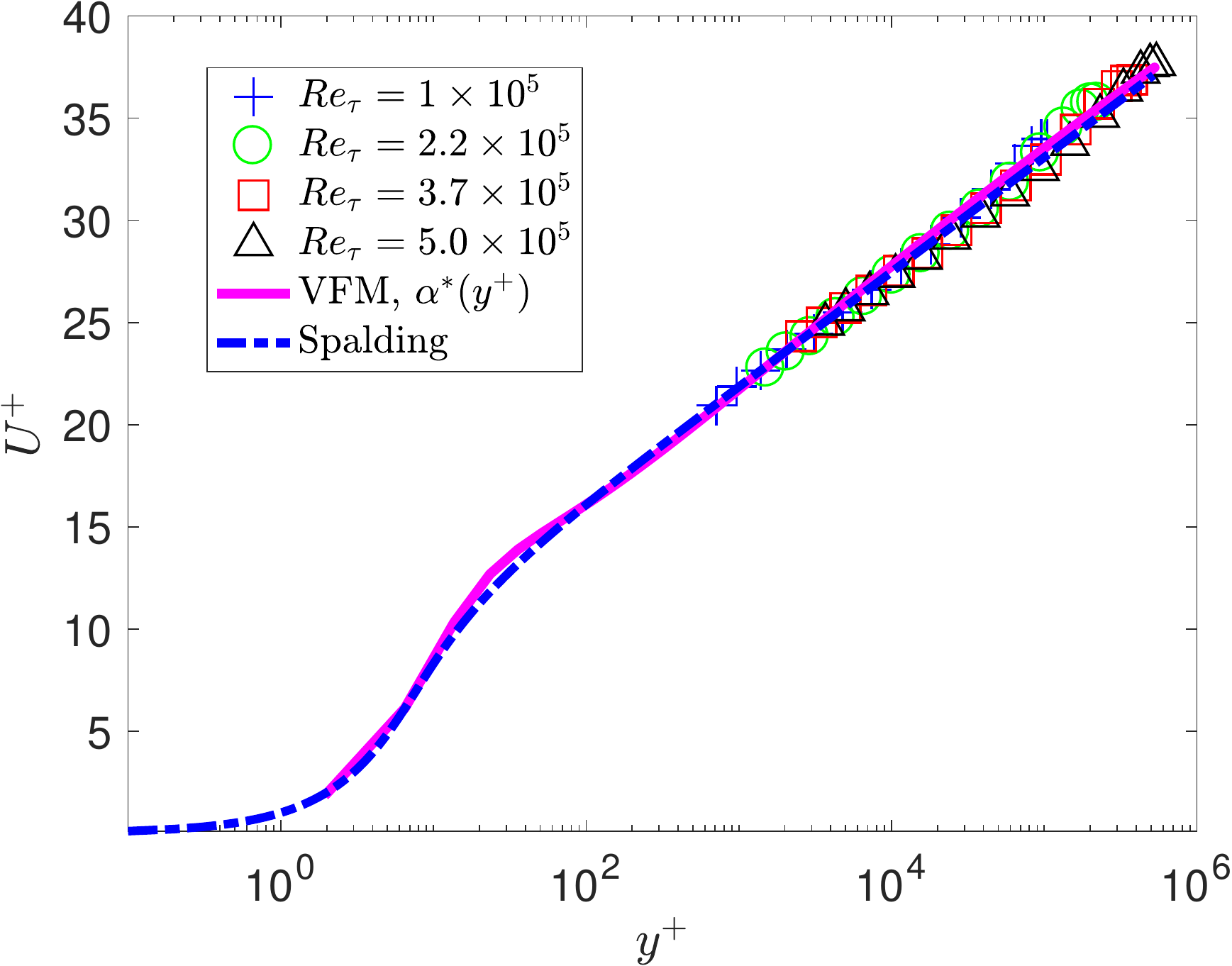}}
\caption{{\bf VFM predictions of mean velocity profile for the superpipe flow form  $Re_\tau=1\times 10^5\ \text{to} \ 5\times10^5$}: (a) velocity profile reconstructed from the experimental data ($\triangle$, \cite{mckeon2002static}), DNS data at $Re_\tau=5200$ ($\square$,\cite{LM2015D}), and the Spalding profile (blue line \cite{white1991viscous}) using  multifidelity Gaussian process regression (M-GPR); (b) velocity profiles  using VFM and the Spalding curve against the experimental data.}
\label{predictive_pipe5e5} 
\end{figure}

Next we consider turbulent {\em pipe flow} and again we test the universal variable fraction order $\alpha(y^+)$ against DNS and experimental data. First, we examine the highest Reynolds number available from the superpipe experiment \cite{zagarola1998mean,mckeon2002static} at $Re_\tau = 5\times10^5$, estimated at $Re_R \approx 3.525\times10^7$ based on the pipe radius $R$.
Given that the experimental data are only available for $y^+>10,000$ we synthesize an {\em entire profile} from
the pipe wall to centerline using multifidelity Gaussian process regression (M-GPR) \cite{kennedy2000predicting} as follows.
We consider as high fidelity data the superpipe data in the outer region together with the highest DNS data for channel flow at $Re_\tau =5200$. We then employ the Spalding curve to provide the low fidelity data and using M-GPR we construct the final profile as shown in Fig. \ref{predictive_pipe5e5}(a). Having this profile and the VFM model transformed in polar coordinates, we can then solve the inverse problem and obtain a new variable fractional order $\alpha(y^+)$. As we show in \cite{Supplementary} this new function is {\em identical} to the formula we derived in Fig. 1.
This finding further confirms the universality of the variable fractional order even at very high Reynolds numbers. Having validated the accuracy of the variable fractional order, we can now solve the forward fractional differential problem to obtain predictions of the entire velocity profiles from $Re_\tau = 10^5$ to $Re_\tau = 5\times 10^5$. 
In Fig. \ref{predictive_pipe5e5}(b) we plot the results and we show that there is excellent agreement with all available data from the superpipe experiment. In addition, in \cite{Supplementary} we present some comparisons with available DNS pipe flow data \cite{wu2008direct} at two different but low Reynolds numbers in \cite{Supplementary}. The agreement is good with DNS data but there is a slight discrepancy at the centerline, which has been discussed in \cite{wu2008direct} as this difference is also present in comparisons with experimental data at comparatively {\em low}  Reynolds numbers. 
Indeed, as pointed out in \cite{wu2008direct}, the universal defect law for pipe flows is not valid for the low Reynolds number range, and this is also in agreement with \cite{zagarola1998mean}
who argued that the lowest $Re_\tau$ for universality is approximately 5,000.


We now discuss an alternative fractional model, where the variable fractional order $\alpha(y)$ is between one and two instead of the VFM we presented where $0< \alpha(y) \le 1$; this model is analogous to VFM and is defined by:
\beq\label{FMII}
\nu_0\frac{\partial^2U}{\partial y^2}+\nu(y)D_y^{\alpha(y)}U=f,\ \forall y\in\Lambda=(0,1], 
\eeq
with $\alpha(0)=2$, and the variable order  $1\leq\alpha(y)\leq2$ an unknown function to be determined by the data.
The scalar coefficient $\nu(y)$ can be also computed from a similar formula as before, i.e.,  
\beq\label{visc2}
\nu(y)=\lim\limits_{y_0\rightarrow\frac{1}{Re_\tau}}\frac{f}{D_y^{\alpha(y)}(U|_{y_0})}.
\eeq
However, unlike the aforementioned VFM, we are unable to obtain an explicit formula for $\nu(y)$, relating it to the Reynolds number as in the first model (i.e., $\alpha(y)\in(0,1]$); instead we can compute it numerically from the DNS data of turbulent channel flow. As shown in \cite{Supplementary}, this alternative fractional model exhibits also a universal scaling if plotted in terms of wall units, with the lowest value of $\alpha(10^5+) \approx 1.3$.

The first fractional model for the Reynolds averaged Navier-Stokes equations was developed by Chen \cite{CW2006FT}, who proposed a fractional Laplacian to  
model the Reynolds stresses and to also account for intermittency \cite{marusic2010wall,Menevau2017} as follows:
\beq
\frac{\partial U}{\partial t}+U\cdot\nabla U=-\nabla p+\nu_0\Delta U-\gamma(-\Delta)^{1/3}U,
\eeq
where $\gamma$ is the turbulent diffusion coefficient. Hence, the effective fractional order in his model is fixed at $\alpha=2/3$. This value is consistent with the
superdiffusion scaling of Richardson for homogeneous turbulence that leads to a $t^3$ scaling for the mean-square-displacement but it is not valid for wall-bounded turbulence where 
anisotropy and the distance from the wall determine the effective rate of turbulent diffusion. Defining a fractional Laplacian in multi-dimensions and in bounded domains is still an open issue in fractional calculus and extending it to variable order is challenging. However, other somewhat equivalent definitions based on {\em tempered} fractional calculus \cite{sabzikar2015tempered} may lead to satisfactory non-local representations as well. As Richardson first noted, the velocity field in the atmosphere shares a number of properties with the Weierstrass function, i.e. it appears to be continuous but non-differentiable, and this make a strong case for fractional modeling of turbulence in the atmosphere but also in wall-bounded flows in engineering applications.


\section*{Acknowledgments}
This work was supported by the OSD/ARO/MURI on ``Fractional PDEs for 
Conservation Laws and Beyond: Theory, Numerics and Applications (W911NF-15-1-0562)".


\section*{List of supplementary materials}
Materials and Methods\\
Figs. 5 to 21 \\
Reference \textit{27-32}


\newpage  
\section*{Supplementary materials}
\subsection*{Materials and Methods} 

{\bf\ \ \ \  Numerical method: Learning the fractional order --}
We assume that we know the mean velocity $U(y)$ (also $U^+(y^+)$) from the DNS data or experimental results. The VFM can be written in the form:
\beq\label{sFM}
\nu(y)D_y^{\alpha(y)}U=f,
\eeq
where $f=\partial P/\partial x-\nu_0\partial^2U/\partial y^2$. The kinematic viscosity $\nu_0$ is typically much smaller than $\nu(y)$ for a high Reynolds number turbulent flow, and the  viscous term is negligible in this case.
Since the fractional order $\alpha(y)$ is unknown in equation \eqref{sFM}, we need to solve a nonlinear problem to obtain $\alpha(y)$. Alternatively, we consider 
the following optimization problem: Given $U$ and $f$, find $\alpha(y)$ that satisfies 
\beq\label{OPR}
J(\alpha(y))=\inf\limits_{\alpha(y)\in S}\|\nu(y)D_y^{\alpha(y)}U-f\|^2,
\eeq
where, $S(\Lambda):=\{0\leq a(y)\leq1,a(y)\in C^0(\Lambda)\}$. If $\alpha^*(y)$ satisfies equation \eqref{sFM}, then we obtain $J(\alpha^*(y))\equiv0$.

Next, we present a numerical method for solving the  optimization problem \eqref{OPR}. The fractional derivative is discretized with the finite-difference method. Then the fractional order $\alpha(y)$ can be solved point-by-point; for each point $y_n=n\Delta y, \Delta y=1/N ,n=1,2,\cdots,N$, we calculate the fractional derivative $D_y^{\alpha(y_n)}U^n$ with the DNS data by using the finite difference method \cite{LXf2007}
\beq
D_y^{\alpha(y_n)}U^n = \frac{1}{\Gamma(2-\alpha(y_n))} \sum\limits_{j=0}^n b^n_j\frac{U^{n+1-j}-U^{n-j}}{\Delta y^{\alpha(y_n)}},
\eeq
where $b^n_j:=(j+1)^{\alpha(y_n)}-j^{\alpha(y_n)}$ and $U^n=U(y_n)$.
The discrete optimization problem can be now written as
\beq\label{DOPR}
J_N(\alpha(y))=\inf\limits_{\alpha(y)\in S} \sum\limits_{n=1}^N\big|\nu(y_n)D_y^{\alpha(y_n)}U^n-f(y_n)\big|^2\Delta y.
\eeq
Here we use $N\approx Re_\tau$ points to solve the above optimization for the channel flow at a given Reynolds number $Re_\tau$. At high Reynolds number ($Re_\tau>10^5$), we use high resolution to compute the fractional variable order when $y^n\ (y^+<200) $ is near the wall, and we use a coarser grid  when $y^n\ (y^+>200)$ far away from the wall. This adaptive strategy  reduces the computational cost for the optimization problems. We will discuss the consequences of rough approximation of the fractional order at the end of this supplementary material. As we showed in the main text, the numerical solutions of the above optimization problem reveal that the fractional order $\alpha(y)$ for all different Reynolds number 
$Re_\tau$ has a universal profile when scaled in wall units.  We fit the fractional order by using these numerical results to obtain the fractional order $\alpha(y^+)$ in wall units as follows
\beq\label{polaw}\ba{c}\dps
\alpha^*(y^+)=\frac{1-{\color{black}\phi(y^+)}}{2}+\frac{{\color{black}\phi(y^+)}+1}{2}{\color{black}a(y^+)},\ea
\eeq
where {\small$\color{black}\phi(y^+)=\tanh(\dps\ln(y^+/9.5)/1.049)$ and ${\color{black}a(y^+)=\dps1/(b+\kappa|\ln(y^+)|^{0.9})}$} with $b=0.855$, $\kappa=0.301$ are constants. 

\begin{figure}[http]
\centering
\subfloat[]{
\includegraphics[height=2.1in]{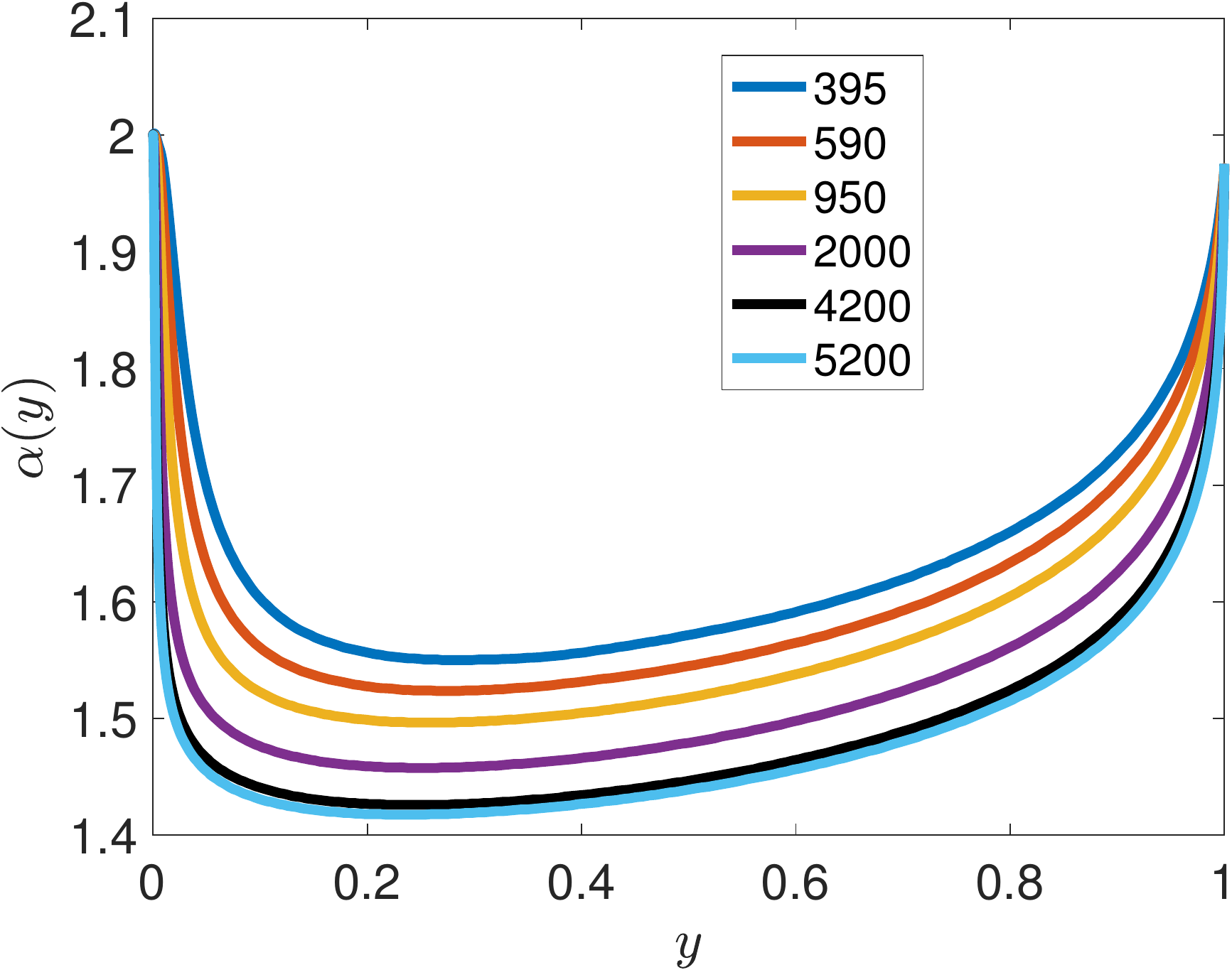}}
\subfloat[]{
\includegraphics[height=2.1in]{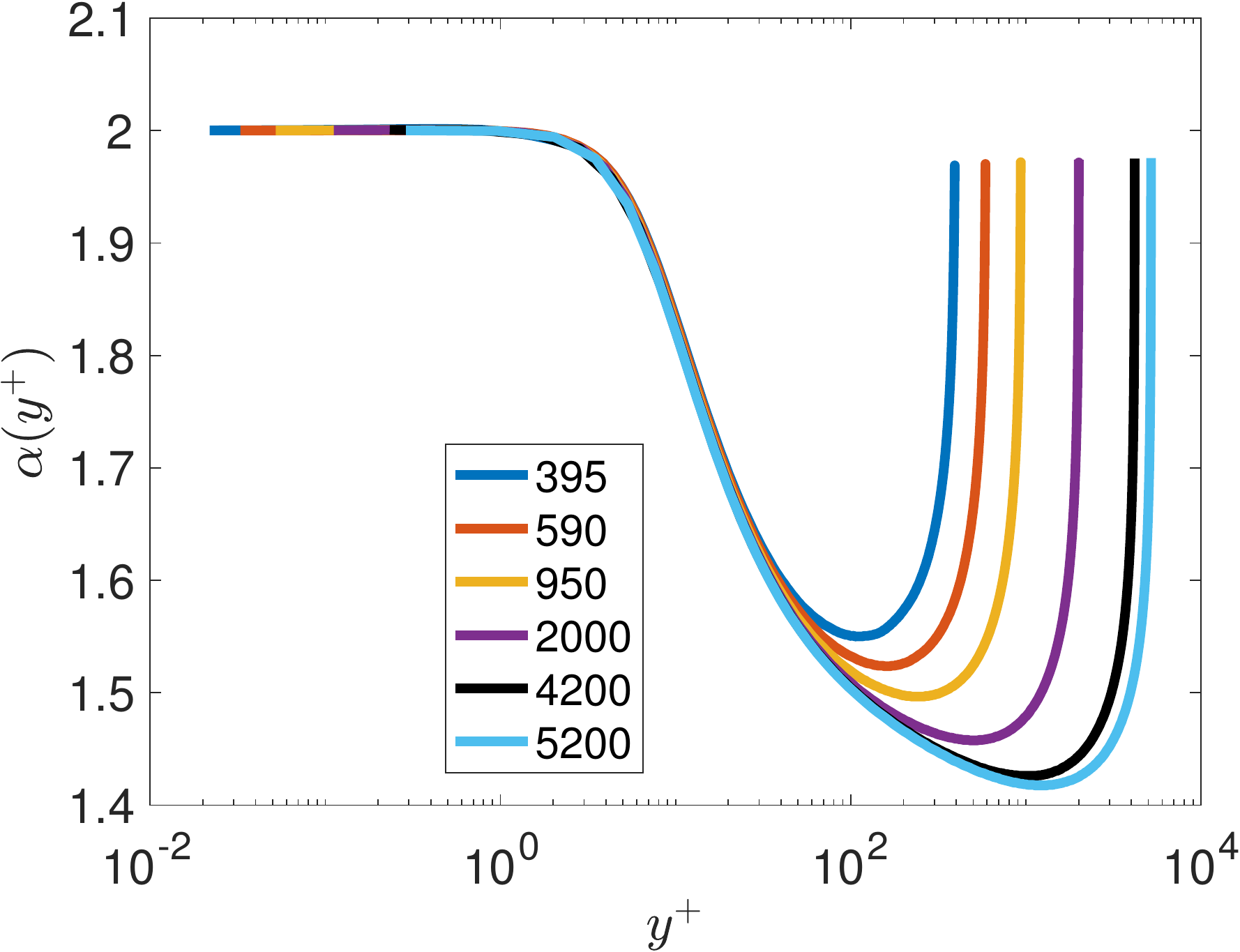}}
\caption{Alternative fractional model in ``conservative'' form.  Learning the fractional order from DNS data of turbulent channel flow.  Fractional order function versus distance from the wall plotted in (a) outer units, (b) wall units.}
\label{fractionalord}
\end{figure}

\vspace{0.1in}
{\bf Alternative Fractional Models --} We will discuss two alternative fractional models in this subsection. First, the variable turbulent model is assumed in the following form,
resembling its integer-counterpart in conservative form: 
\beq
D_y(-\nu(y)D_y^{\alpha(y)-1}U)=f,
\eeq
where the eddy viscosity is defined as $\nu(y)=\Gamma(3-\alpha(y))Re_\tau^{1-\alpha(y)}V/u_\tau$. Integrating by parts we can obtain:
\beq
\nu(y)D_y^{\alpha(y)-1}U=(1-y)f.\eeq 
Fig. \ref{fractionalord} presents the variable fractional orders we obtained from DNS data for the turbulent channel flow. The numerical results show that the variable fractional order of this model does not exhibit any universality.

\begin{figure}[H]
\centering
\subfloat[]{
\includegraphics[height=2.1in]{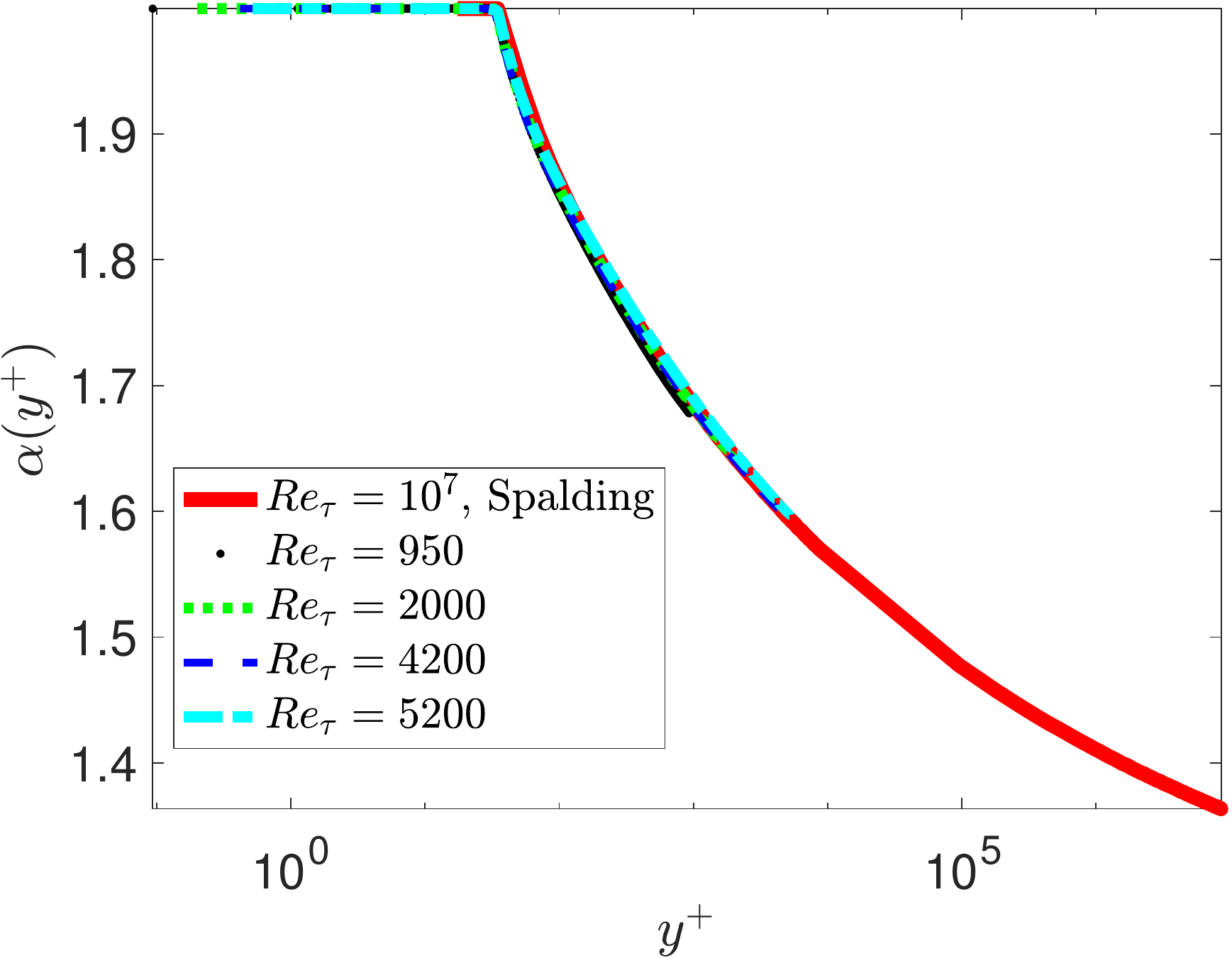}}
\subfloat[]{
\includegraphics[height=2.1in]{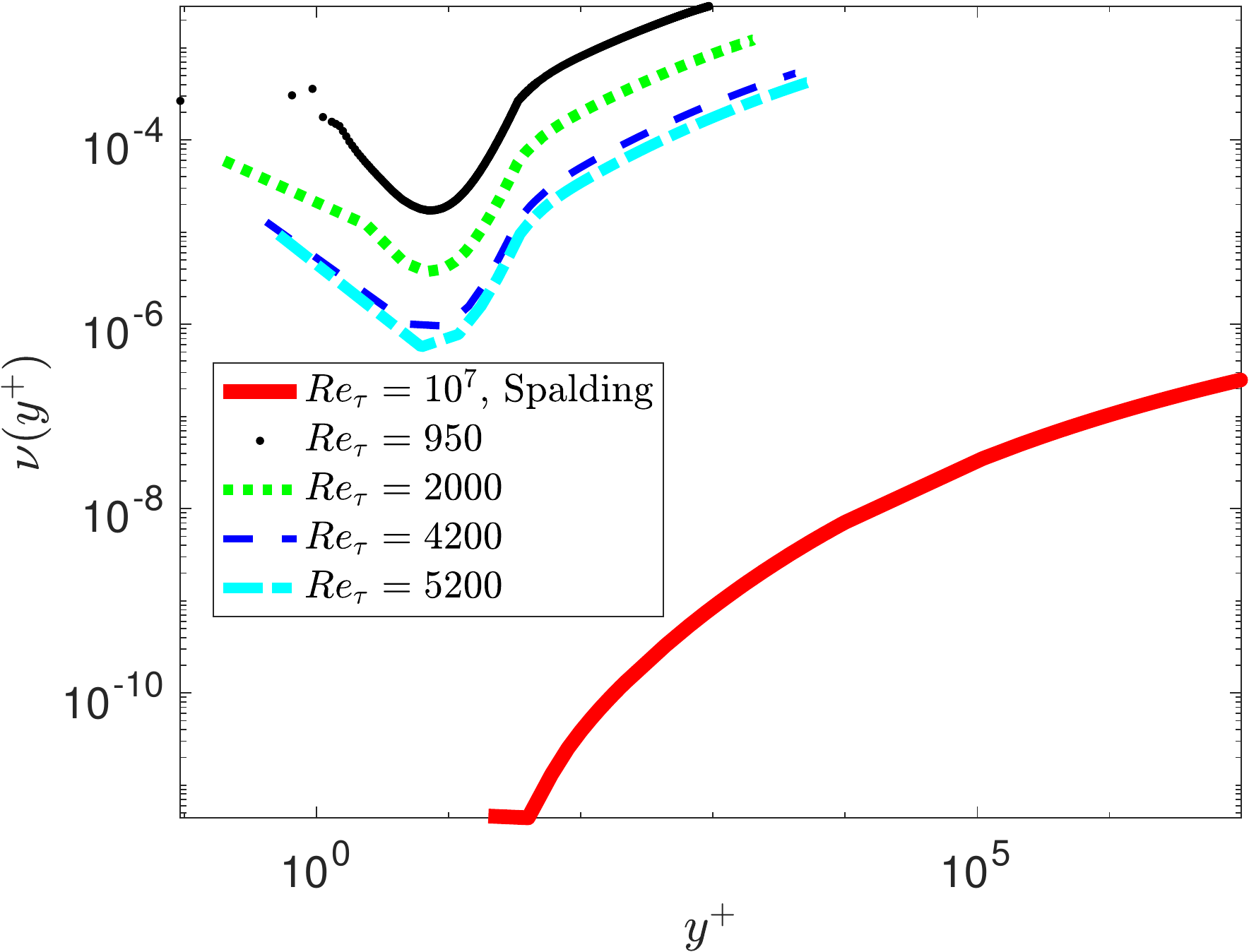}}\\ 
\caption{Alternative  fractional model with $1 \le \alpha(y) \le 2$. The numerical fractional orders are computed based on DNS data for turbulent channel flow 
at $Re_\tau=950,2000,4200,5200$: (a) plots of the fractional orders $\alpha(y^+)$ in wall units; (b) corresponding eddy viscosity coefficients.}
\label{alpha_y2}
\end{figure}

In the main text we also discussed an alternative model, where the variable fractional order $\alpha(0)=2$, and $1\leq\alpha(y)\leq2$.
Unlike the VFM, here the eddy viscosity coefficient is computed numerically from 
\beq\label{viscss2}\nu(y)=\lim\limits_{y_0\rightarrow\frac{1}{Re_\tau}}\frac{f}{D_y^{\alpha(y)}(U|_{y_0})}.\eeq
Here we use the channel flow DNS data, and we show again that the fractional order has a universal profile in wall units.  The discretized scheme of this alternative model is the same  as the variable model VFM. The numerical results are presented in Fig. \ref{alpha_y2}, and indeed we observe that the variable fractional order obeys the universal scaling for Reynolds numbers up to $Re_\tau=10^{7}$. At high Reynolds number, we employed the mean velocity profile from the Spalding
formula at $Re_\tau=10^{7}$.

\begin{figure}[H]
\centering
\subfloat[]{
\includegraphics[height=2.2in]{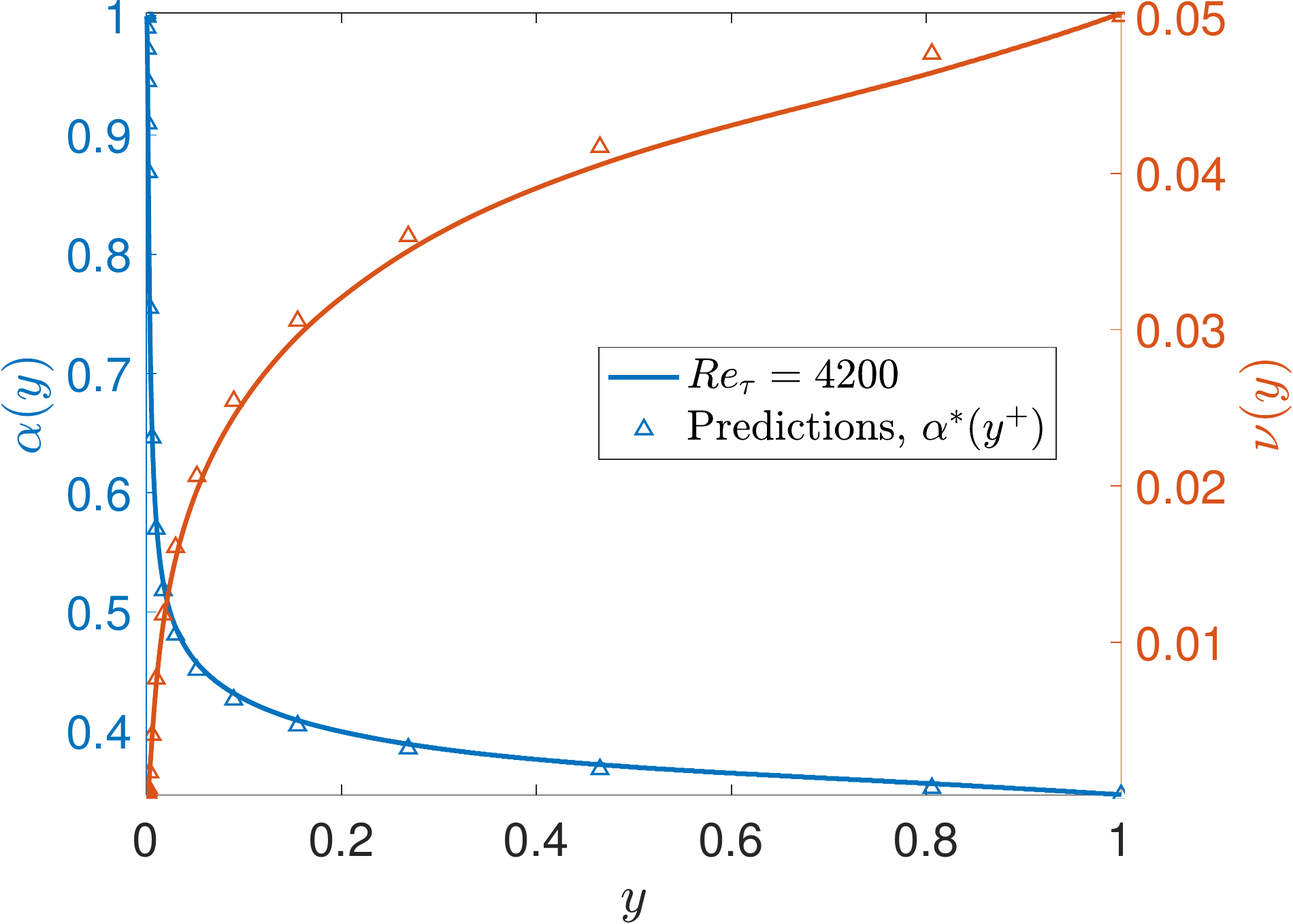}}
\subfloat[]{
\includegraphics[height=2.2in]{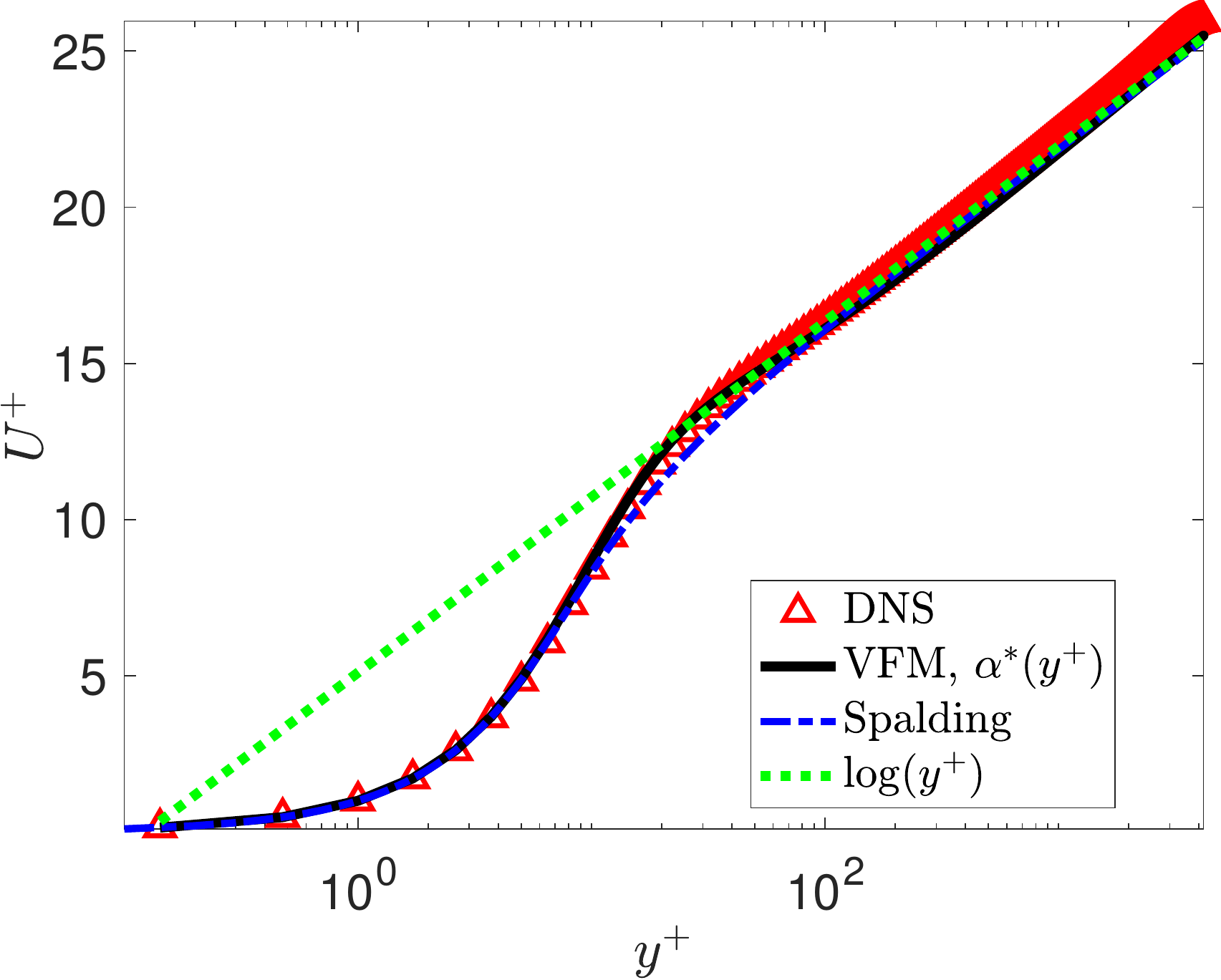}}
\caption{Model predictions for the turbulent channel flow at $Re_\tau=4200$: (a) Solid line ($-$) represents the numerical solution of 
the optimization problem
 and triangle symbols ($\triangle$) represent equation \eqref{polaw}. The blue line represents the fractional order $\alpha(y)$ and the red line the
 eddy viscosity coefficient. This Reynolds number $Re_\tau=4200$ is not included in the training of the model; (b) mean velocity obtained by VFM corresponding to the fractional order $\alpha^*(y^+)$ from the left plot.}
\label{predictive_4200}
\end{figure}

\begin{figure}[H]
\centering
\includegraphics[height=2.2in]{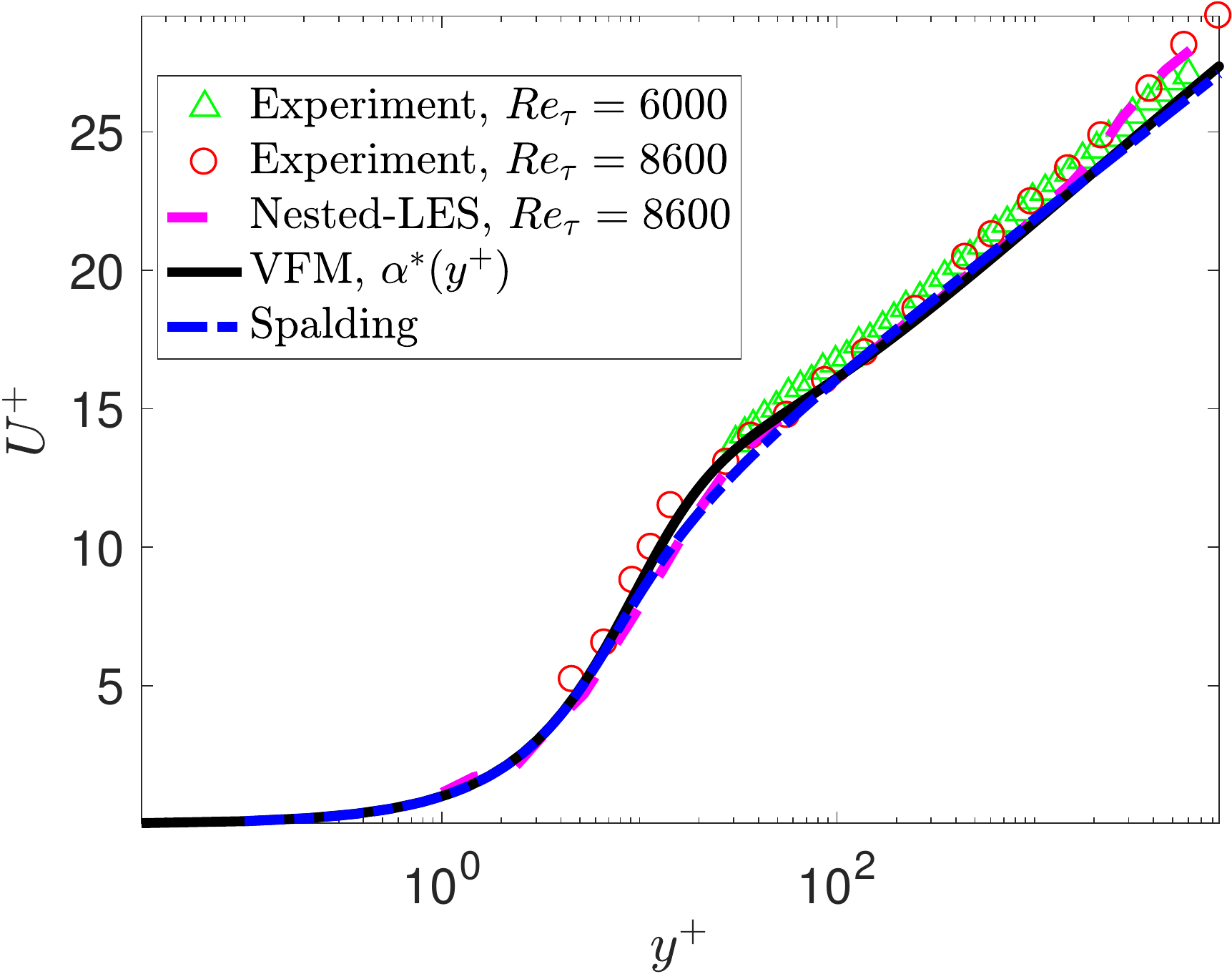}
\caption{Profiles of mean velocity for turbulent channel flow at $Re_\tau=6000,8600$: Triangle symbol ($\triangle$) represents experimental data from \cite{schultz2013reynolds}, circle symbol ($\circ$) represents experimental data from \cite{C1969Turb}, Solid line ($-$) represents VFM profile, Dash line ($--$) represents LES results \cite{tang2016nested}.}
\label{Re8600mean}
\end{figure}

\vspace{0.1in}
{\bf Turbulent channel flow -- }
We test the predictive model for $Re_\tau=[4200,6000,8600]$, i.e., at Reynolds numbers not used in the training of the model. The fractional order $\alpha(y^+)$ is defined in \eqref{polaw}. 
Since there is no available DNS data at $Re_\tau>5200$, we use experimental data in cases of $Re_\tau=6000$ \cite{schultz2013reynolds} and  $Re_\tau=8600$ \cite{C1969Turb} for comparison.
We also include the turbulent channel flow results obtained by nested-LES  \cite{tang2016nested}.  Fig. \ref{predictive_4200} and \ref{Re8600mean} show that the mean velocity profiles predicted by VFM exhibit the correct behavior throughout the channel for Reynolds number up to $Re_\tau=8600$, including the correct slope in the logarithmic layer, and agreement with DNS and experimental data in the wake region for all $Re_\tau=[4200,6000,8600]$.

\begin{figure}[H]
\centering
\subfloat[]{
\includegraphics[height=2.2in]{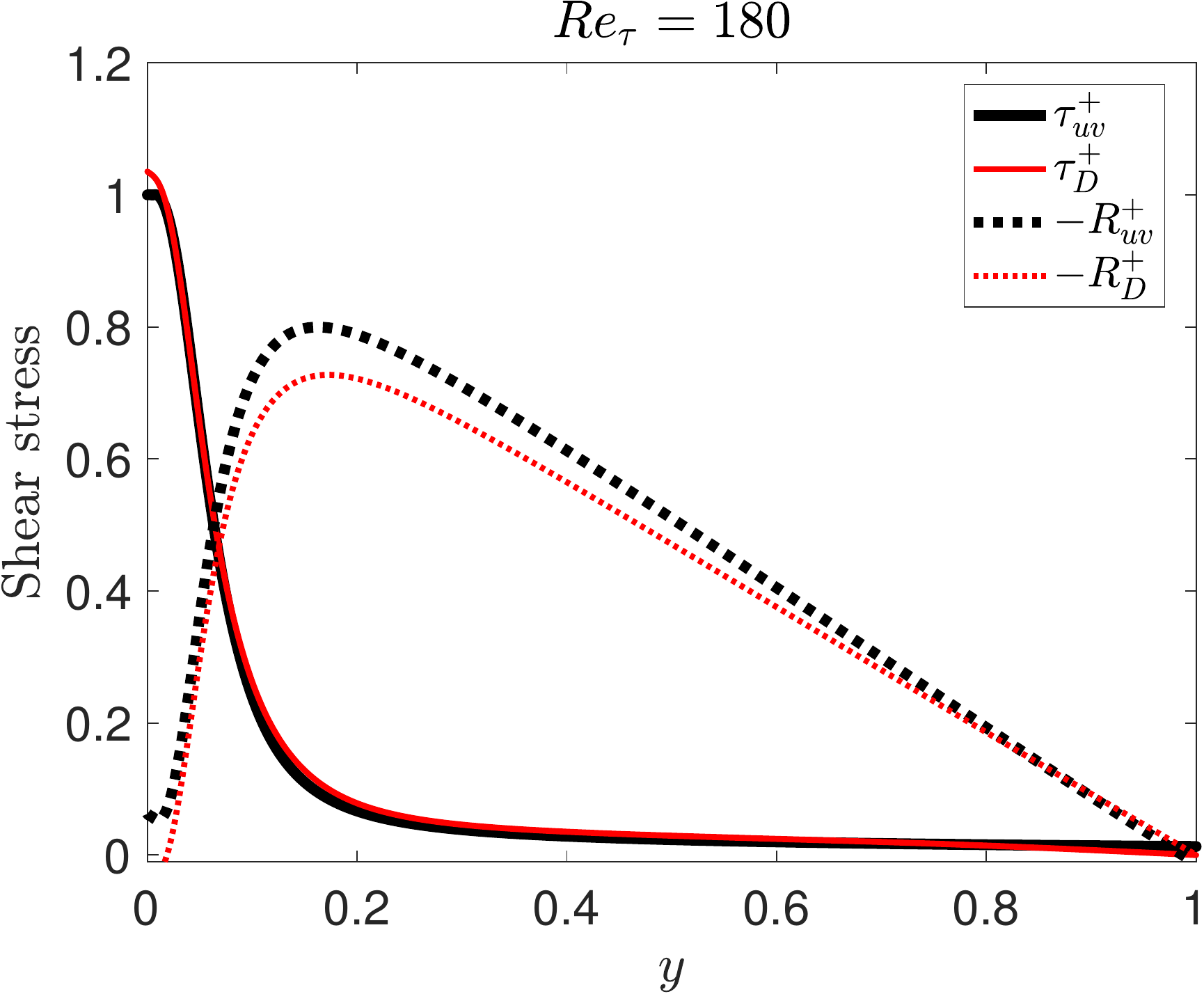}}
\subfloat[]{
\includegraphics[height=2.2in]{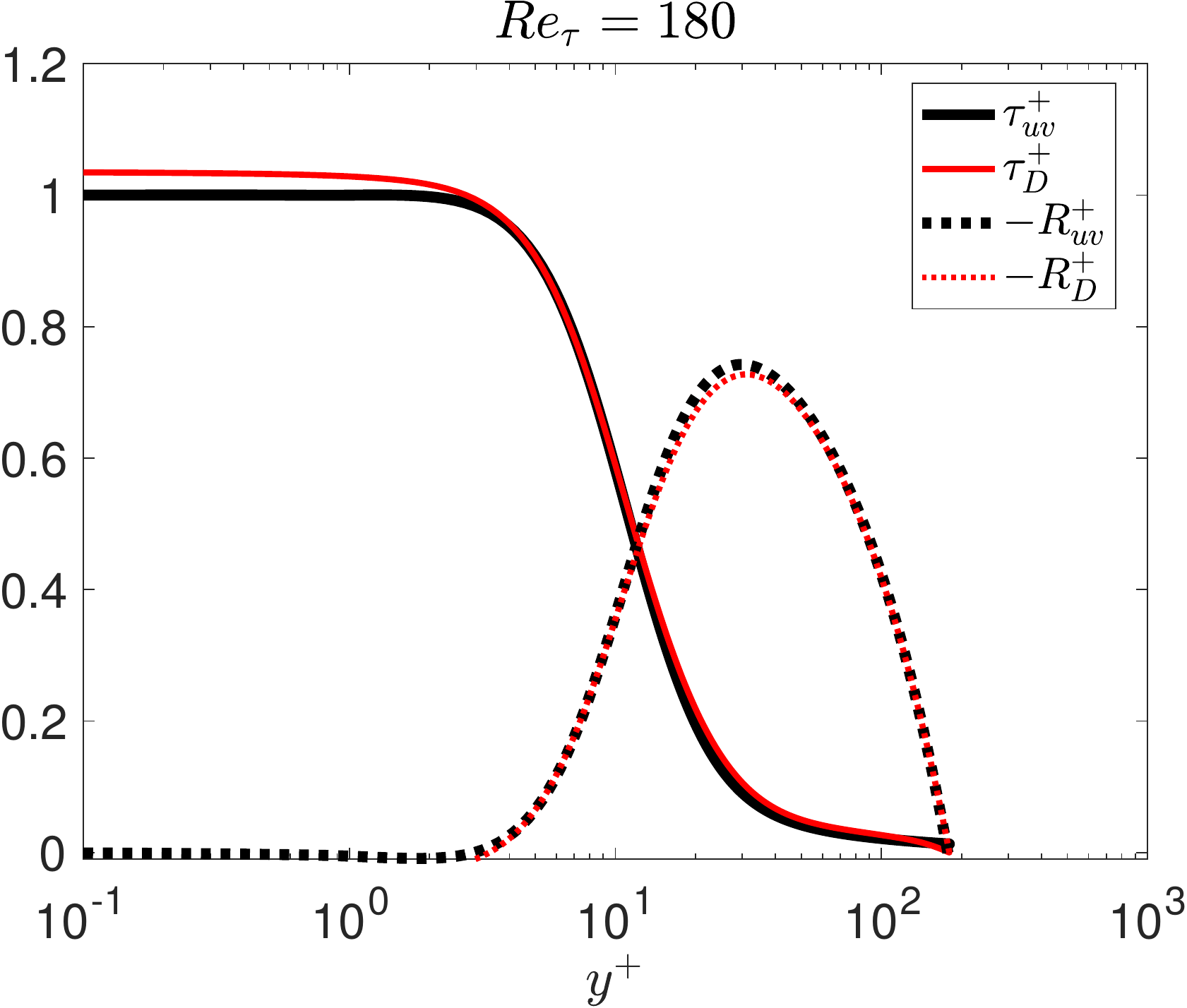}}
\caption{Accurate prediction of the shear stress at $Re_\tau=180$ in outer units and wall units. Here $\tau_{uv}$ denotes the wall shear stress predicted by VFM (black-solid line) and  $\tau_D$ is the  corresponding profile from DNS data. $-R_{uv}$ (black-dash line) denotes the Reynolds shear stress predicted by VFM, and $-R_{D}$ is the corresponding profile from DNS data.}
\label{profile_stress}
\end{figure}

\begin{figure}[H]
\centering
\subfloat[]{
\includegraphics[height=2.2in]{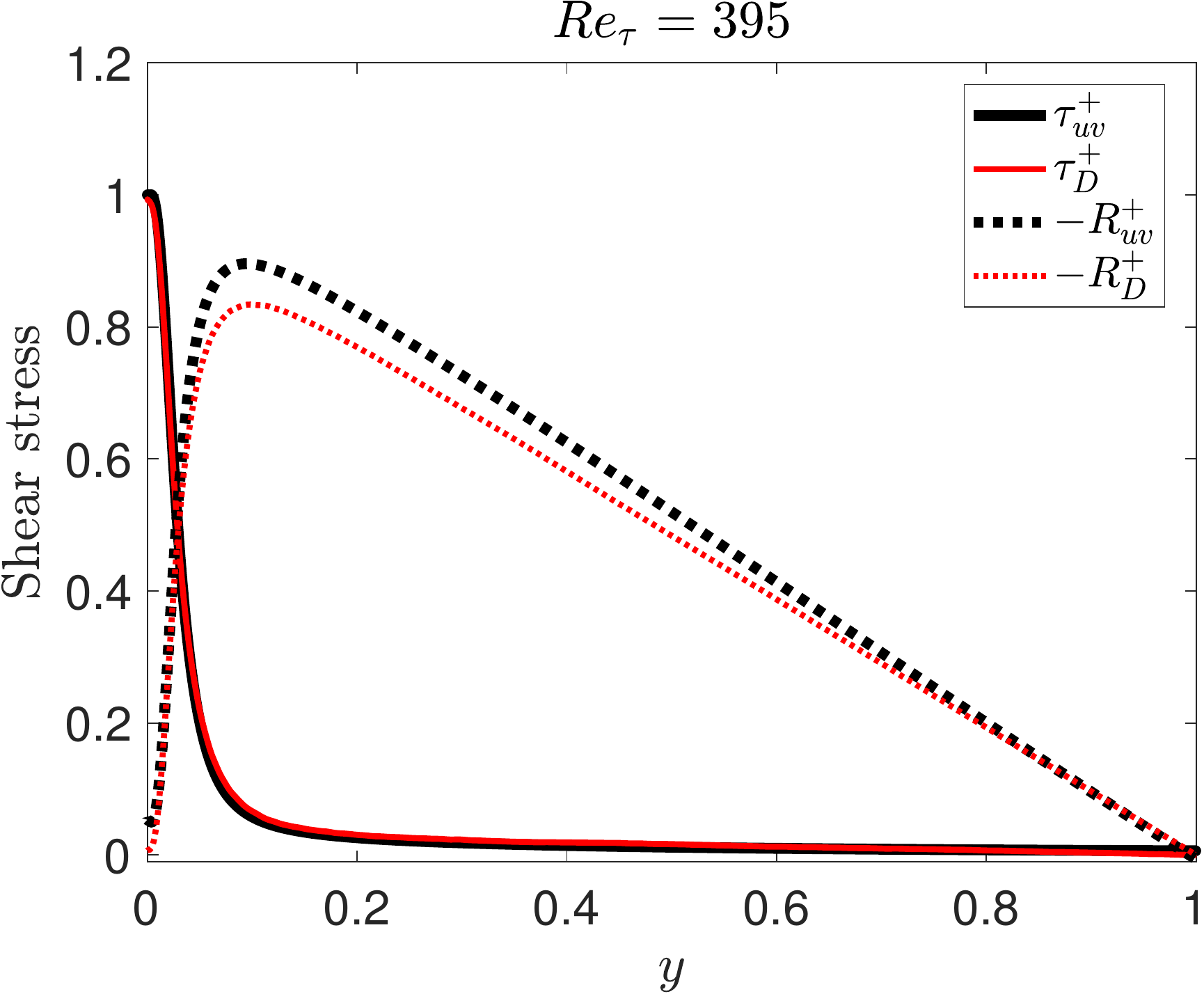}}
\subfloat[]{
\includegraphics[height=2.2in]{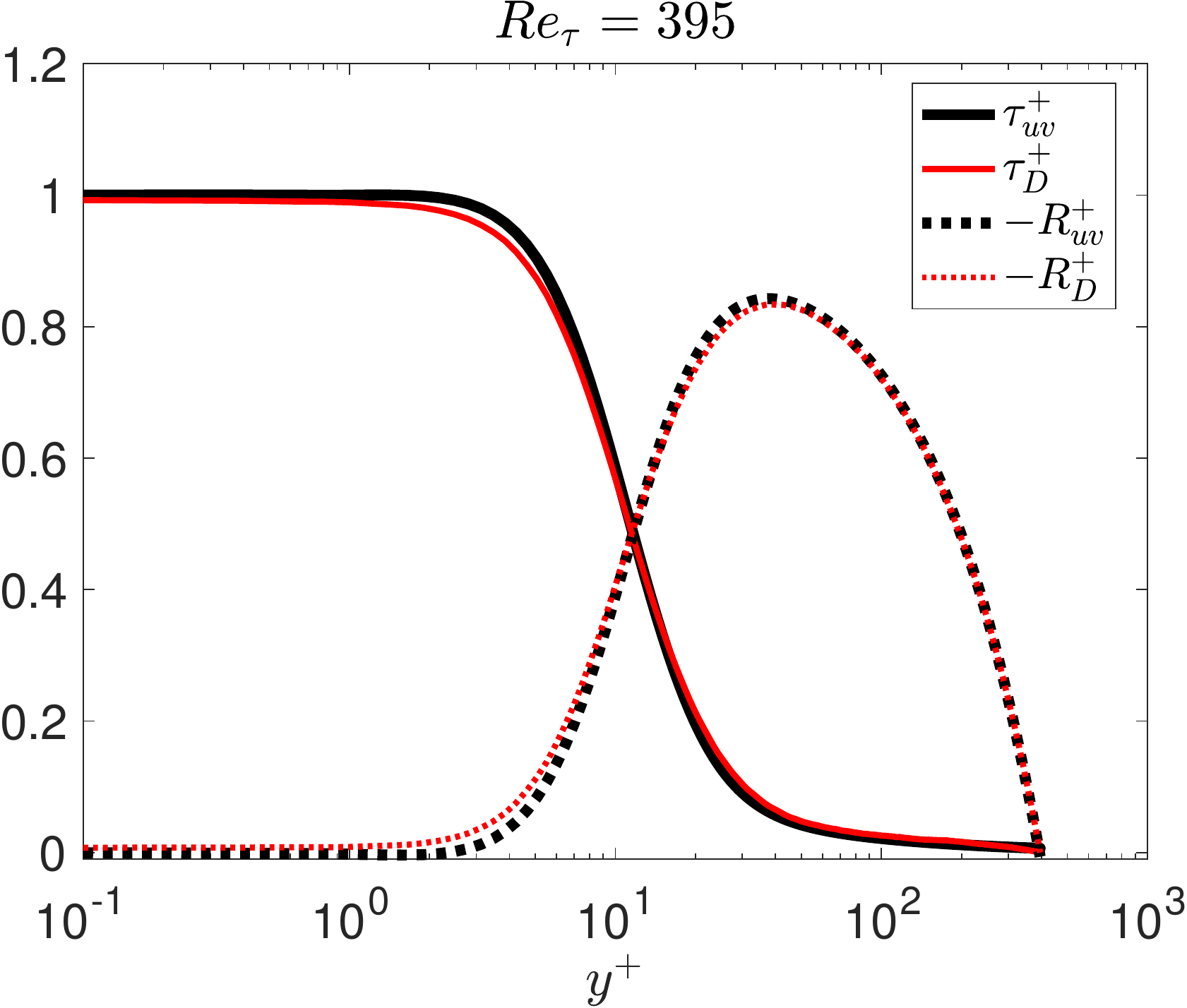}}
\caption{Accurate prediction of the shear stress at $Re_\tau=395$ in outer units and wall units. Caption: see Fig. \ref{profile_stress}.}
\label{profile_stress395}
\end{figure}

\begin{figure}[H]
\centering
\subfloat[]{
\includegraphics[height=2.2in]{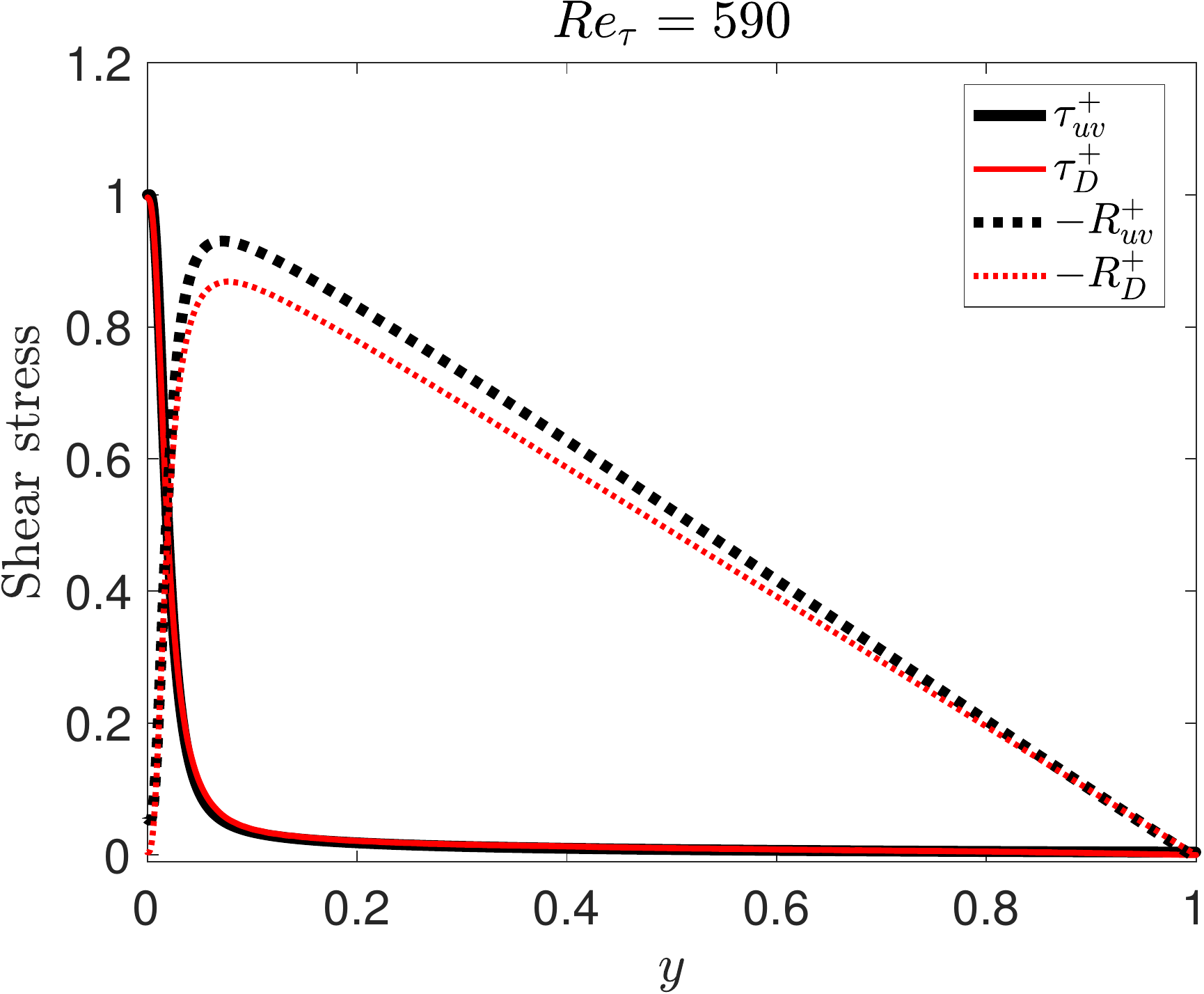}}
\subfloat[]{
\includegraphics[height=2.2in]{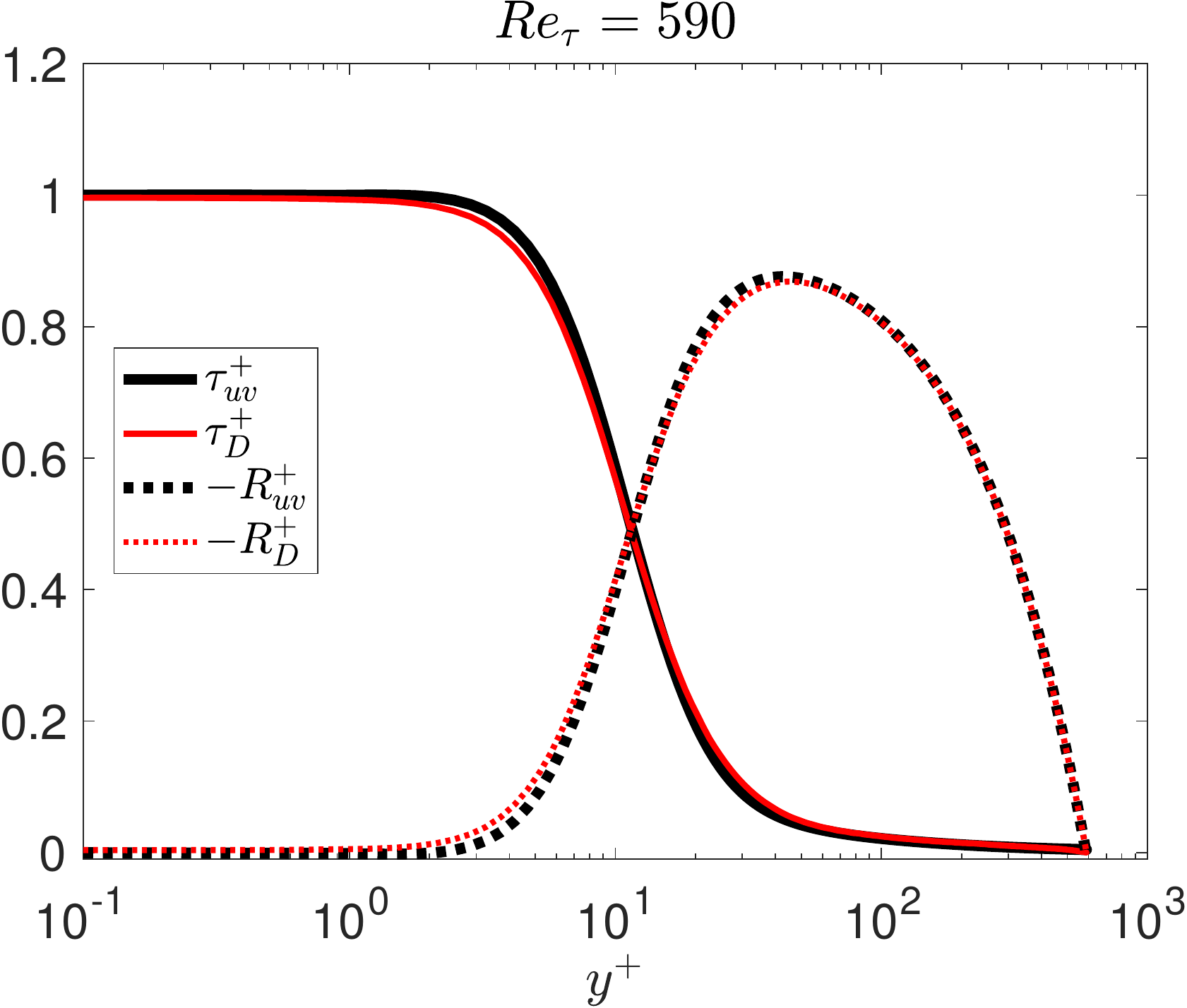}}
\caption{Accurate prediction of the shear stress at $Re_\tau=590$ in outer units and wall units. Caption: see Fig. \ref{profile_stress}.}
\label{profile_stress590}
\end{figure}

\begin{figure}[H]
\centering
\subfloat[]{
\includegraphics[height=2.2in]{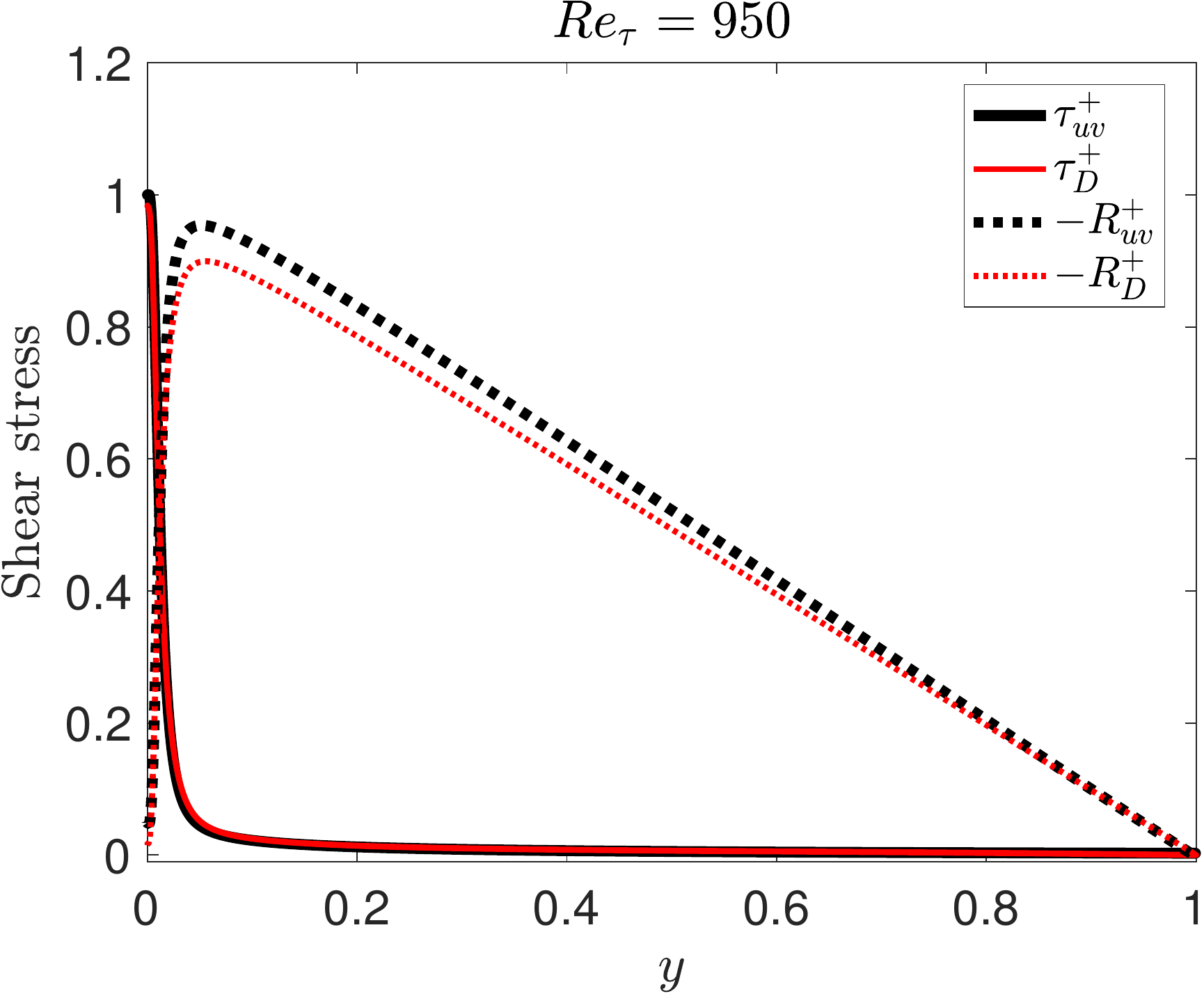}}
\subfloat[]{
\includegraphics[height=2.2in]{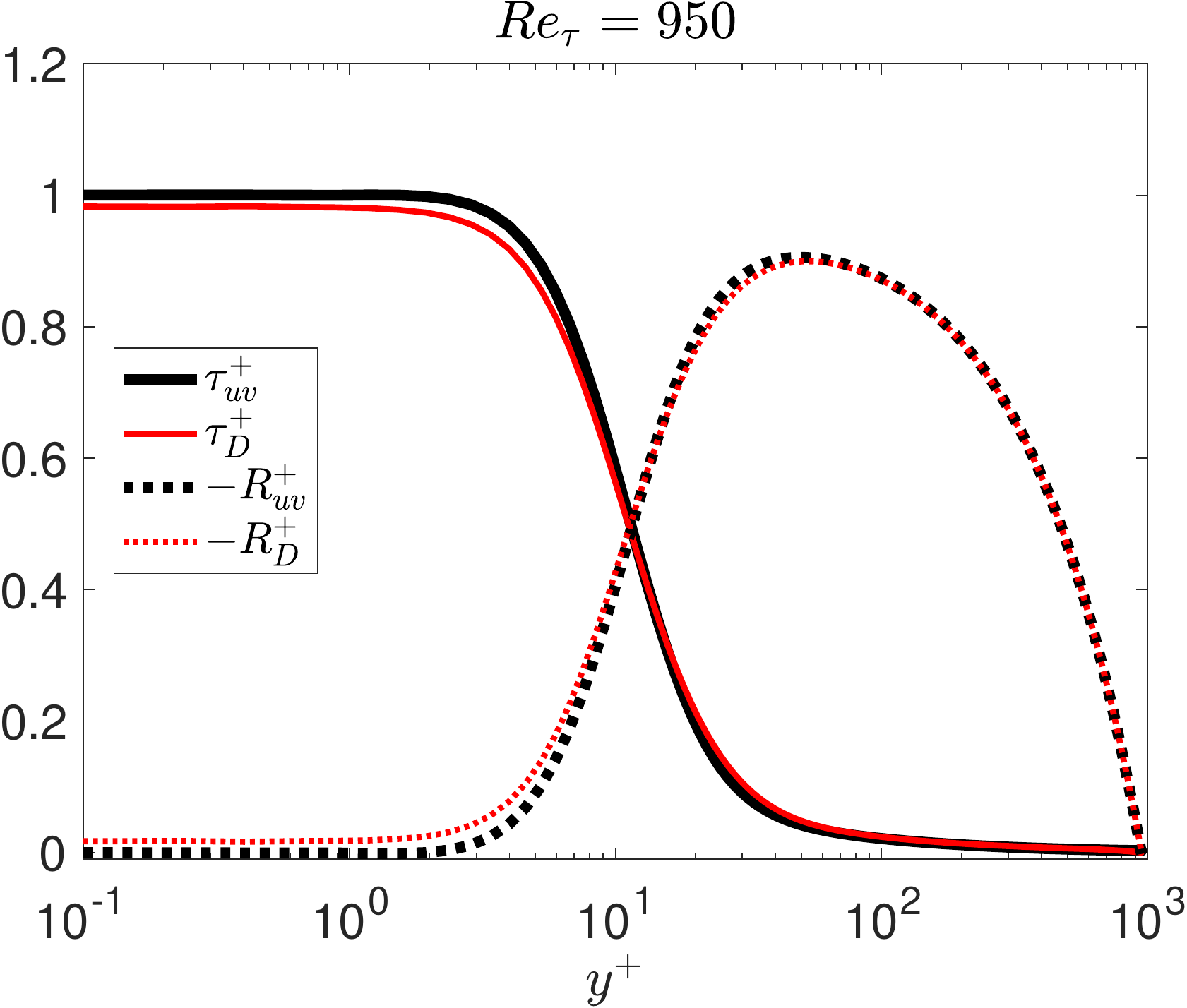}}
\caption{Accurate prediction of the shear stress at $Re_\tau=950$ in outer units and wall units. Caption: see Fig. \ref{profile_stress}.}
\label{profile_stress950}
\end{figure}

\begin{figure}[H]
\centering
\subfloat[]{
\includegraphics[height=2.2in]{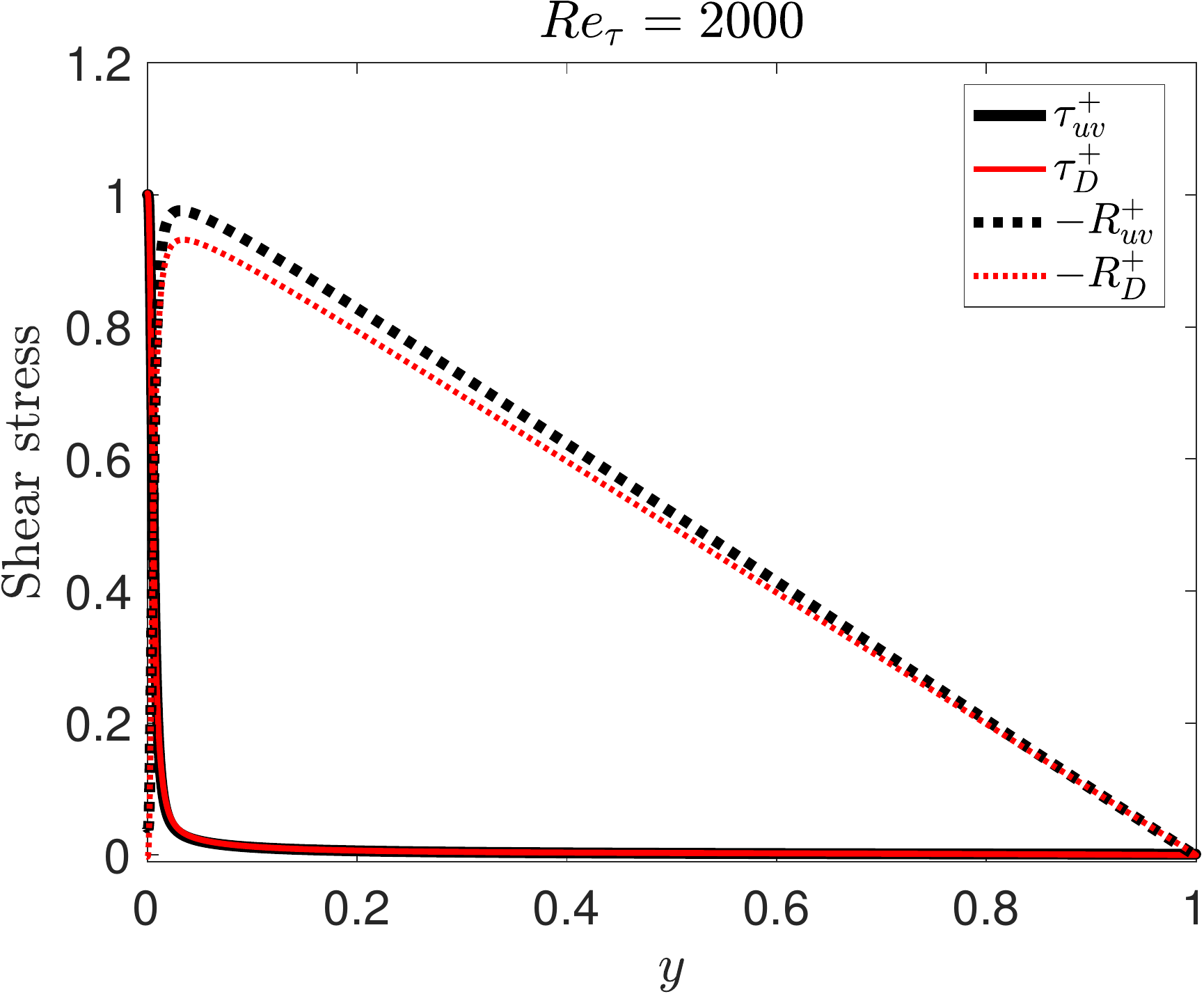}}
\subfloat[]{
\includegraphics[height=2.2in]{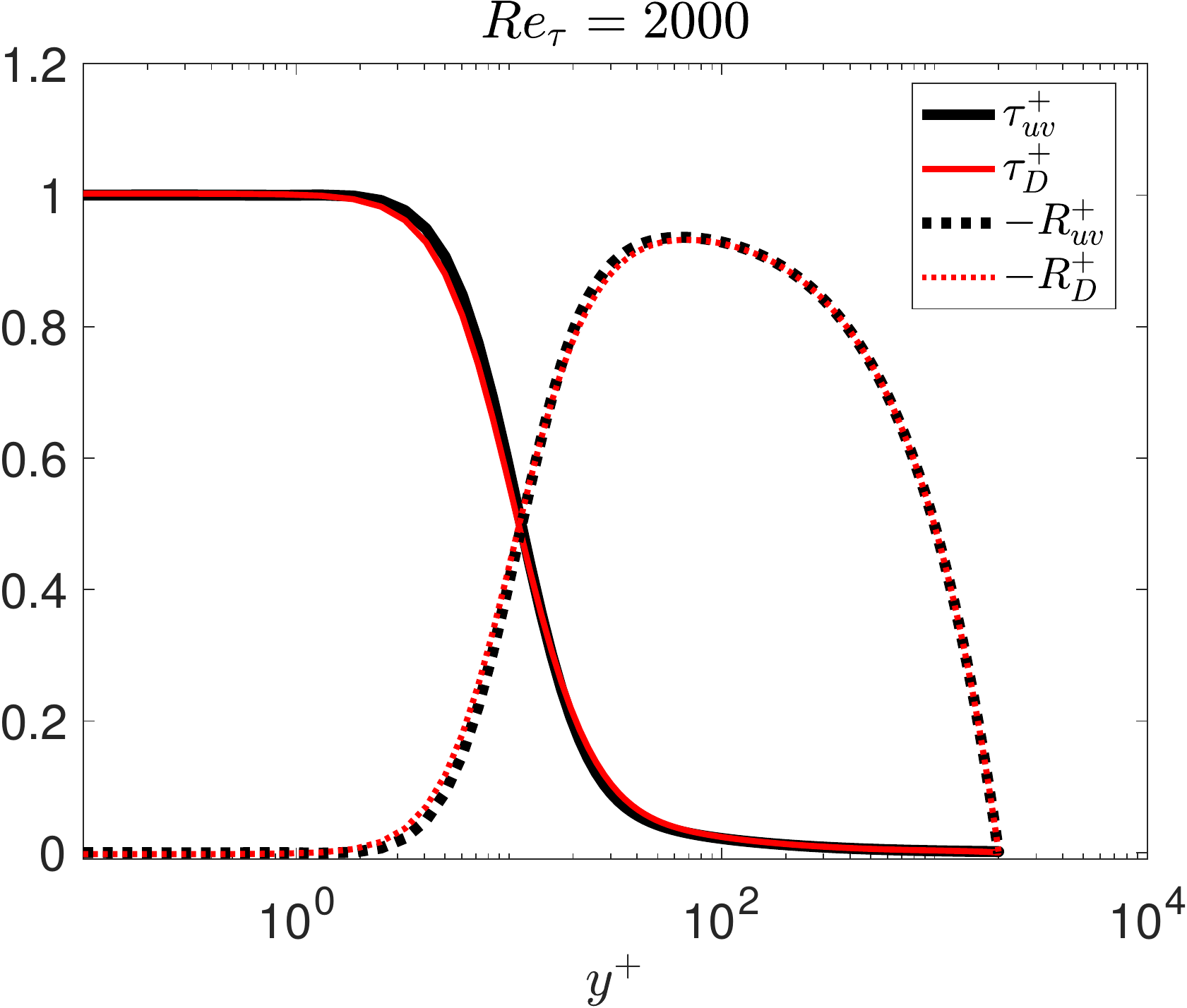}}
\caption{Accurate prediction of the shear stress at $Re_\tau=2000$ in outer units and wall units. Caption: see Fig. \ref{profile_stress}.}
\label{profile_stress2000}
\end{figure}

\begin{figure}[H]
\centering
\subfloat[]{
\includegraphics[height=2.2in]{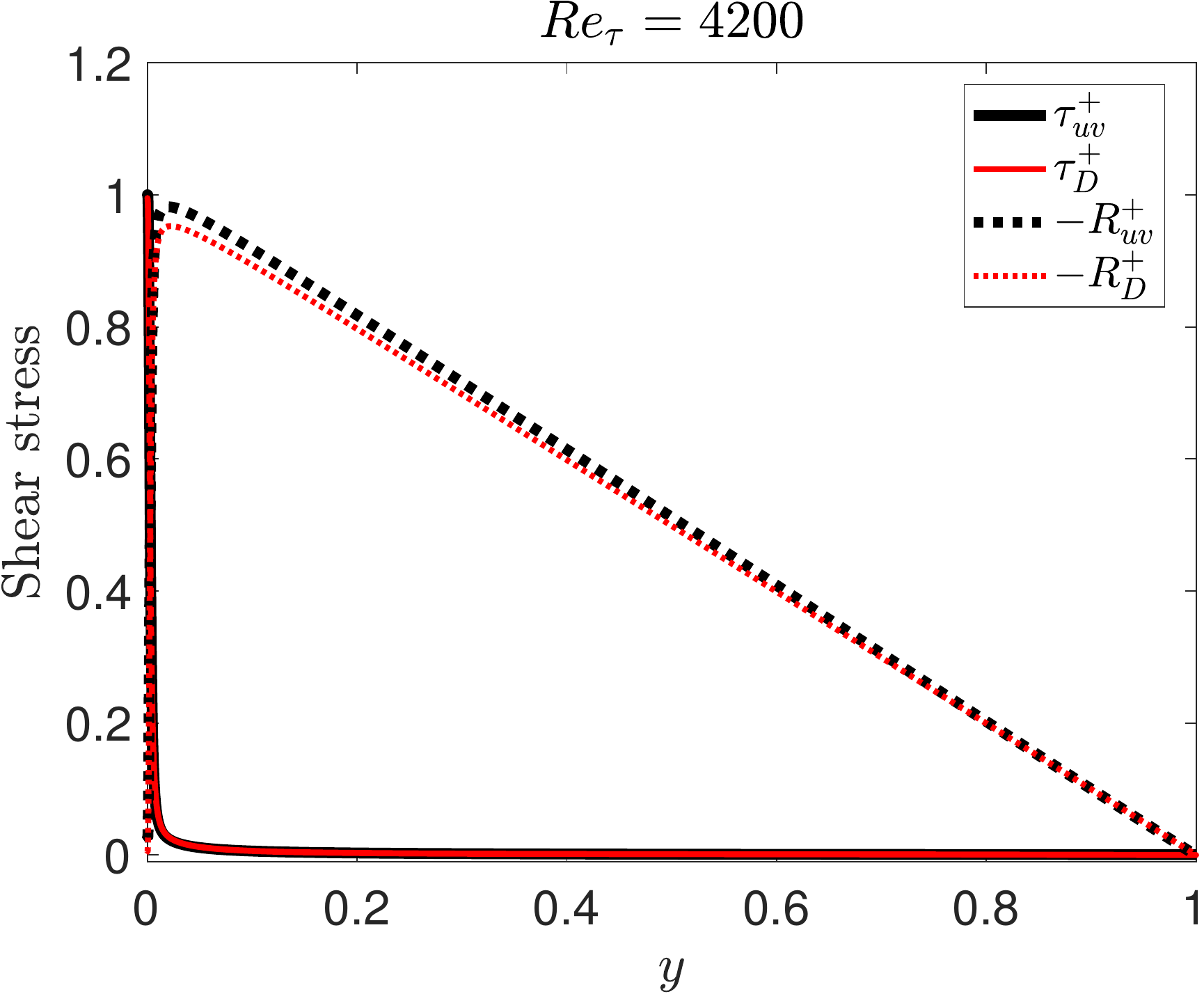}}
\subfloat[]{
\includegraphics[height=2.2in]{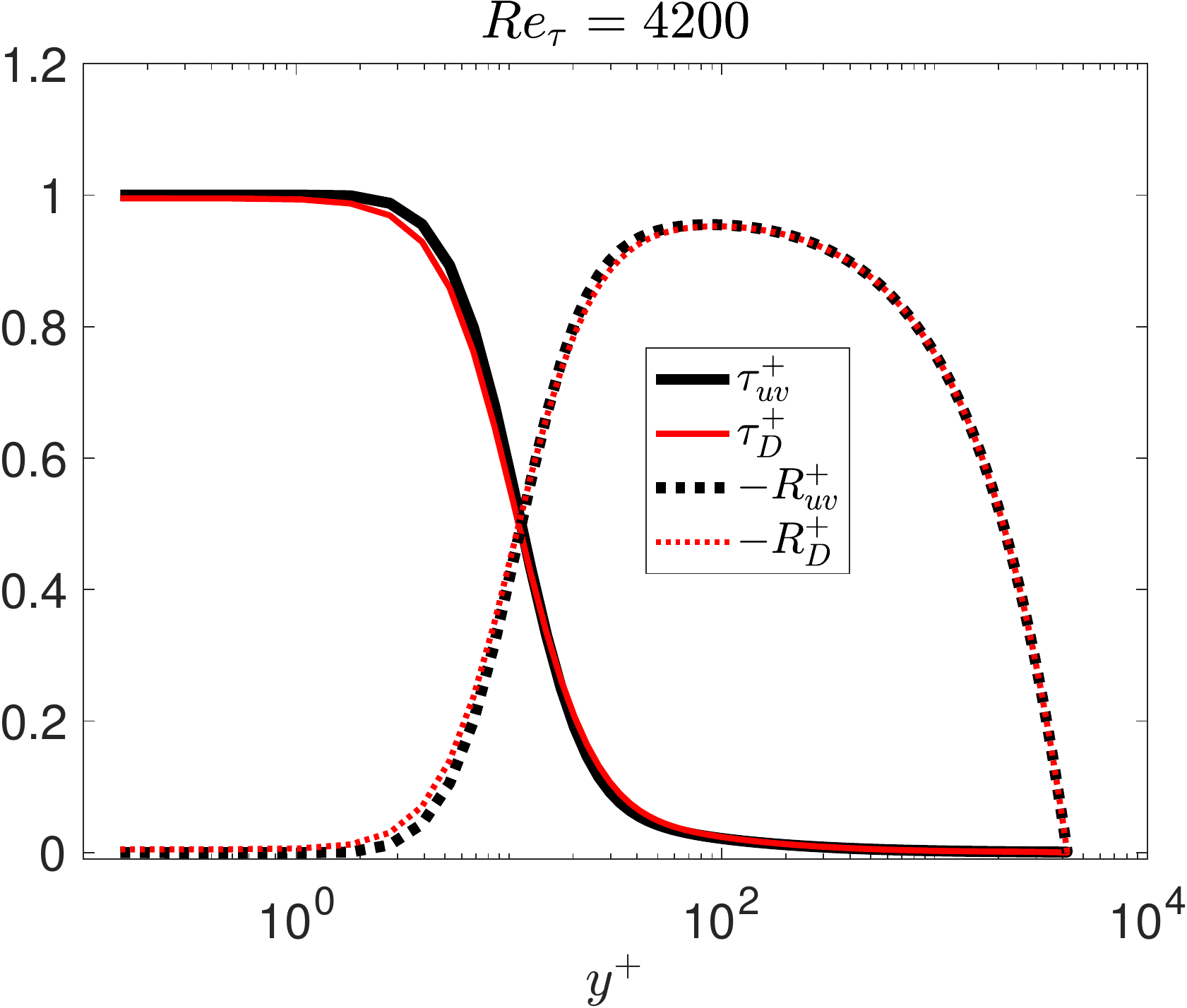}}
\caption{Accurate prediction of the shear stress at $Re_\tau=4200$ in outer units and wall units. Caption: see Fig. \ref{profile_stress}.}
\label{profile_stress4000}
\end{figure}

\vspace{0.1in}
{\bf Reynolds Stresses --} We use the simplified one-dimensional equation 
\be\label{stress1}
\frac{\partial }{\partial y}\big(\tau_{uv}-R_{uv}\big)
=\nu(y){}D_y^{\alpha(y)}U=\frac{\partial P}{\partial x},\ y\in(0,1),\ee
where the $R_{uv}$ denotes the Reynolds stress $R_{uv}=\overline{u'v'}$, $\tau_{uv}$ denotes the viscous shear stress $\tau_{uv} = \nu_0\partial U/\partial y$, and $U$ is the mean velocity, which is the solution to the above fractional equation \eqref{stress1}. Then, we obtain the Reynolds stresses by integration, 
\be\label{Restre2}
-R_{uv} = \int_y^1\nu(s){}D_s^{\alpha(s)}Uds-\tau_{uv}.
\ee
We can compare the predicted Reynolds stresses with their counterparts, $R_D$, from DNS data for turbulent channel flow, and also using the corresponding viscous shear stress denoted by  $\tau_{D}=\mu\partial U_D/\partial y$, where $U_D$ denotes the mean velocity from the DNS data base. Figs. \ref{profile_stress}-\ref{profile_stress4000} plot the predicted and DNS profiles for all Reynolds numbers  from $Re_\tau=180\ \text{to}\ 4200$; very good agreement is observed. 

\vspace{0.1in}
{\bf Turbulent Couette Flow --}
In reference  \cite{Epps2017turbulence}, the authors proposed the double-log
profile to predict the mean velocity for the Couette flow as follows 
\be\label{lppre}
U(y)=\frac{1}{2}-\frac{1}{2}\frac{\ln\big((d+y)/(d+1-y)\big)}{\ln\big(d/(d+1)\big)},
\ee
where $d$ is a small number $(d\ll1)$ that represents a viscous sublayer or roughness height. The non-dimensional boundary conditions are $U(0)=0$ and $U(1)=1$.

\begin{figure}[H]
\centering
\includegraphics[height=2.2in]{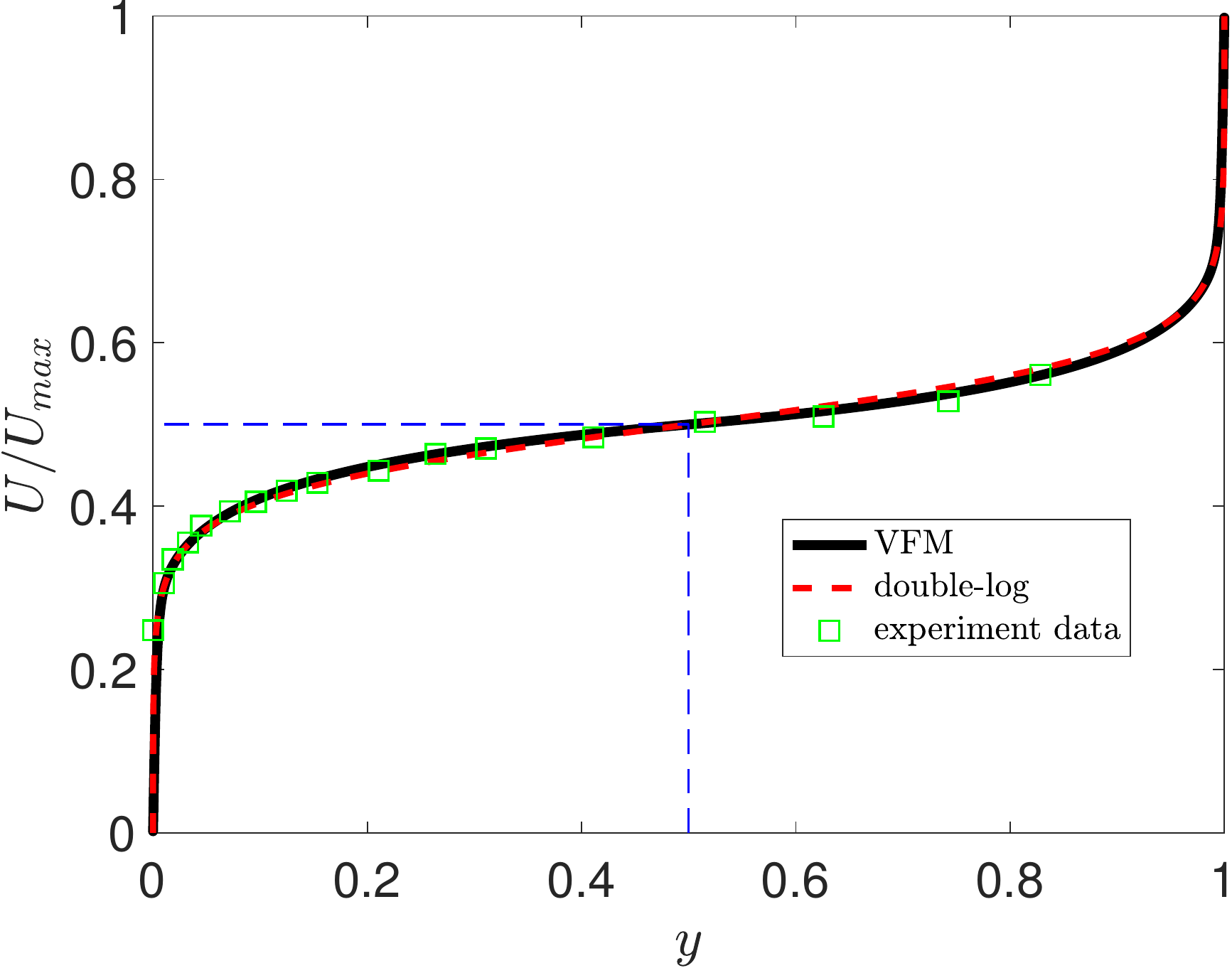}
\caption{Turbulent Couette flow - numerical results for $Re=16500$: ``-" VFM predictions at $Re_\tau=1650$, ``- -" best fit of the double-log profile in equation \eqref{lppre} with $d=1.06\times10^{-5}$, ``$\square$" experimental data from \cite{robertson1970turbulence}.}
\label{profile_y}
\end{figure}

\begin{figure}[H]
\centering
\includegraphics[height=2.2in]{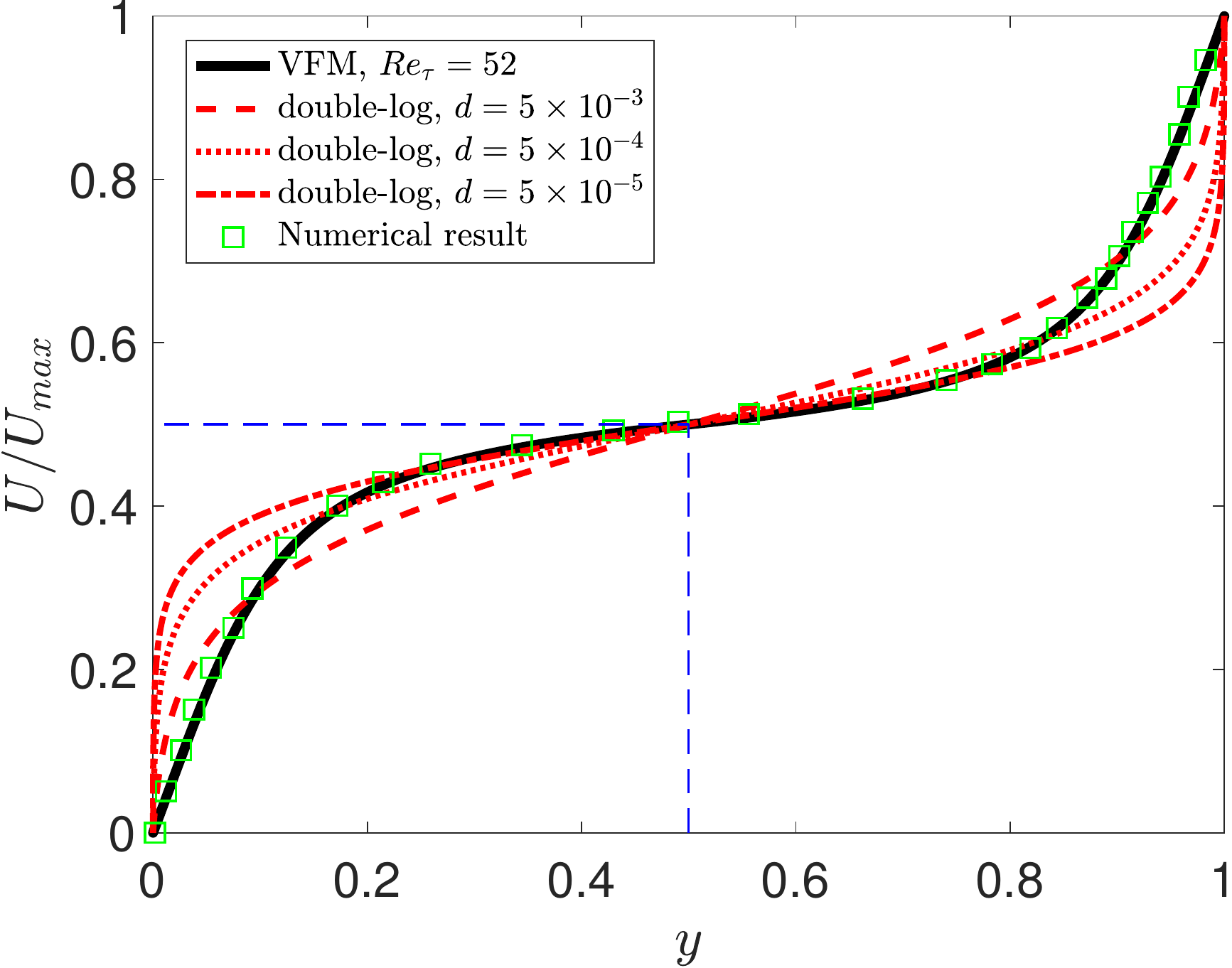}
\caption{Turbulent Couette flow at $Re_\tau=52$: ``-" VFM predictions at $Re_\tau=52$, the dashed lines represent 
the double log profiles in equation \eqref{lppre} with different coefficients $d=5\times10^{-3},5\times10^{-4},5\times10^{-5}$, ``$\square$" represents the numerical result in reference
\cite{liu2003turbulent} (Fig. 1 (a)).}
\label{profile_Re52}
\end{figure}

\begin{figure}[H]
\centering
\subfloat[]{
\includegraphics[height=2.2in]{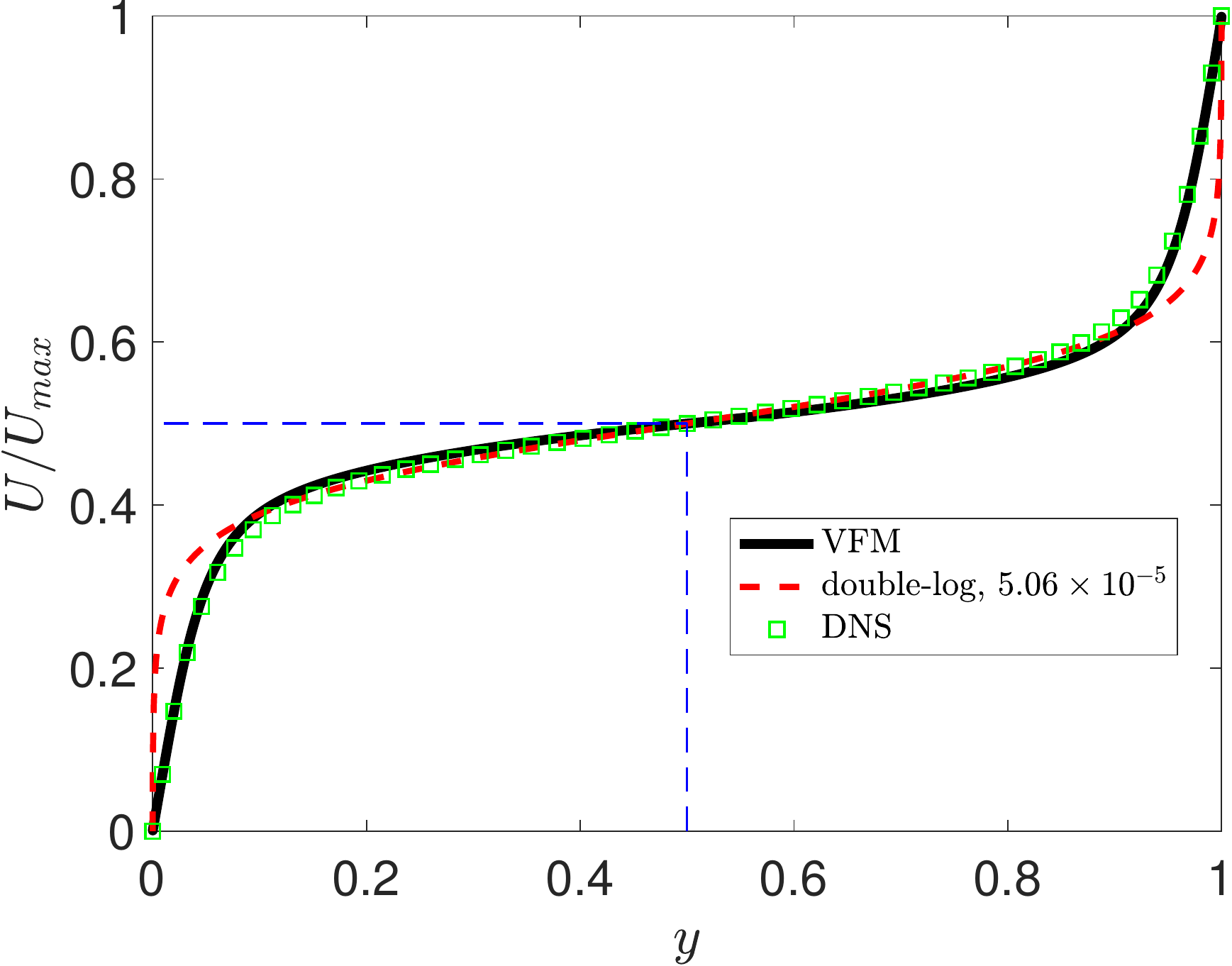}}
\subfloat[]{
\includegraphics[height=2.2in]{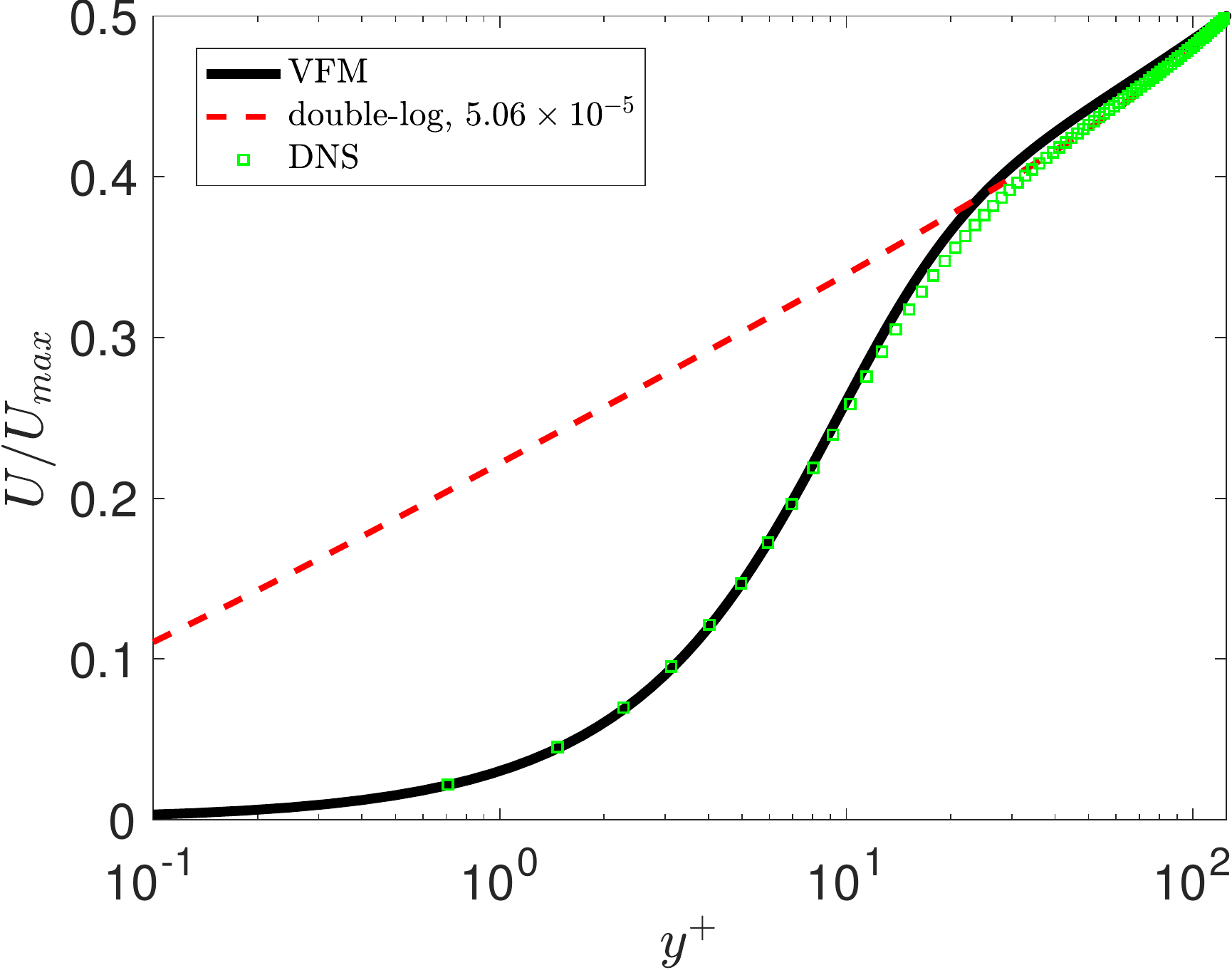}}
\caption{Turbulent Couette flow at $Re_\tau=125$: (a) ``-" VFM predictions, ``- -" best fit of the double-log profile in equation \eqref{lppre} with $d=5.06\times10^{-5}$,  ``$\square$" DNS data  \cite{avsarkisov2014turbulent}; (b) Wall units scaling for the mean velocity profiles.}
\label{profile_y125}
\end{figure}

\begin{figure}[H]
\centering
\subfloat[]{
\includegraphics[height=2.2in]{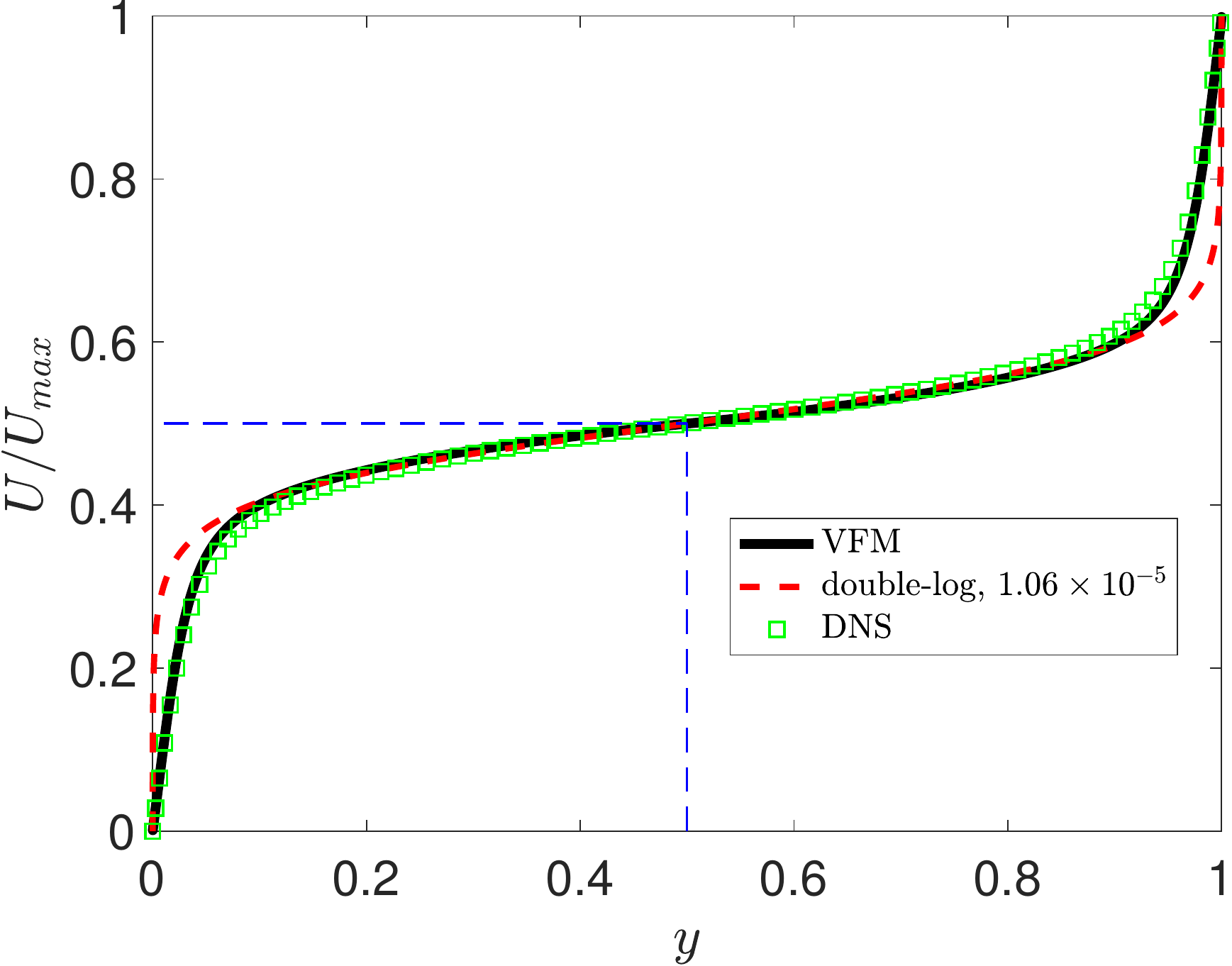}}
\subfloat[]{
\includegraphics[height=2.2in]{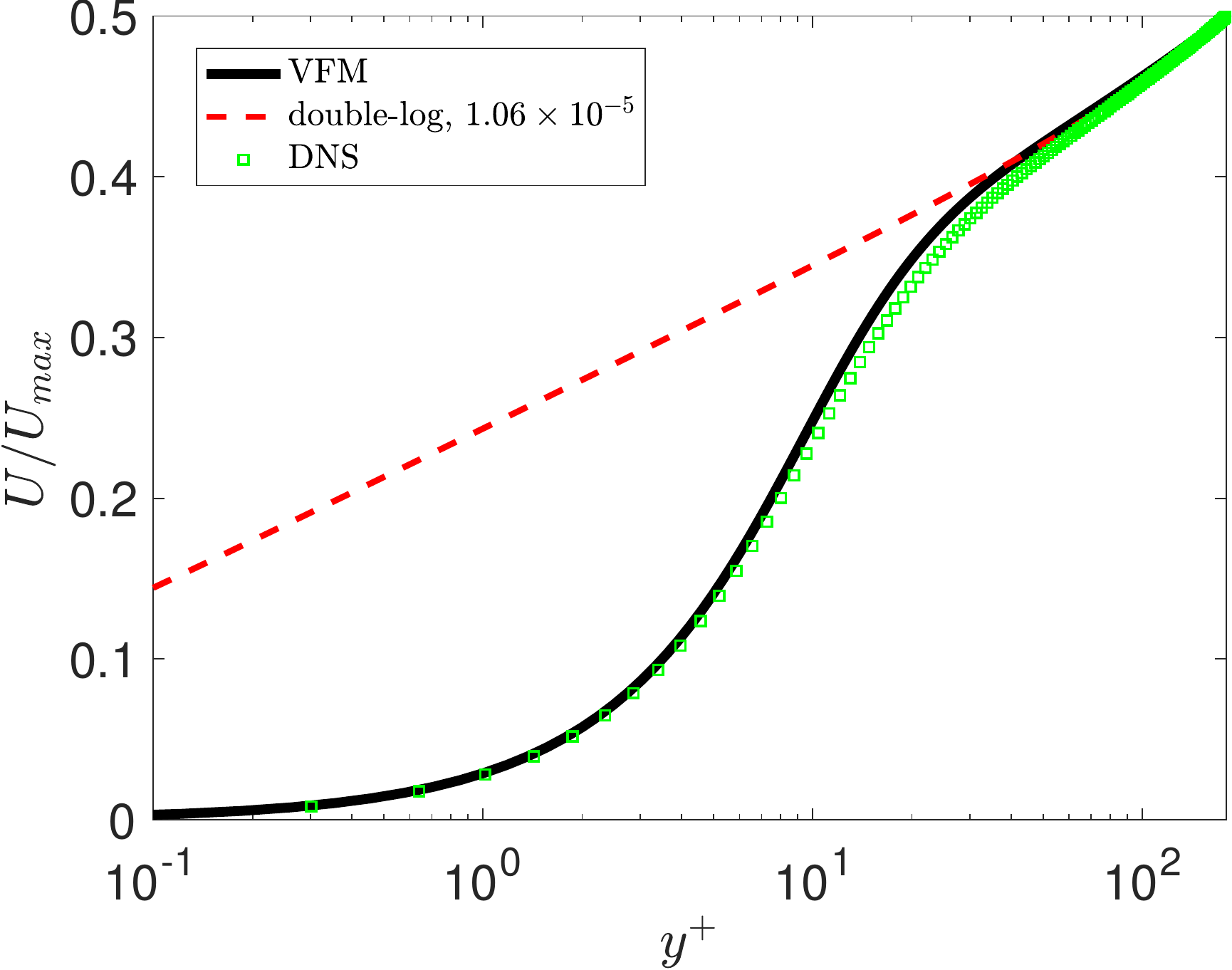}}
\caption{Turbulent Couette flow at $Re_\tau=180$: (a) ``-" VFM predictions, ``- -" best fit of the double-log profile in equation  \eqref{lppre} with $d=1.06\times10^{-5}$, ``$\square$" DNS data  from \cite{avsarkisov2014turbulent}; (b) Wall units scaling for the mean velocity profiles.}
\label{profile_y180}
\end{figure}

\begin{figure}[H]
\centering
\subfloat[]{
\includegraphics[height=2.2in]{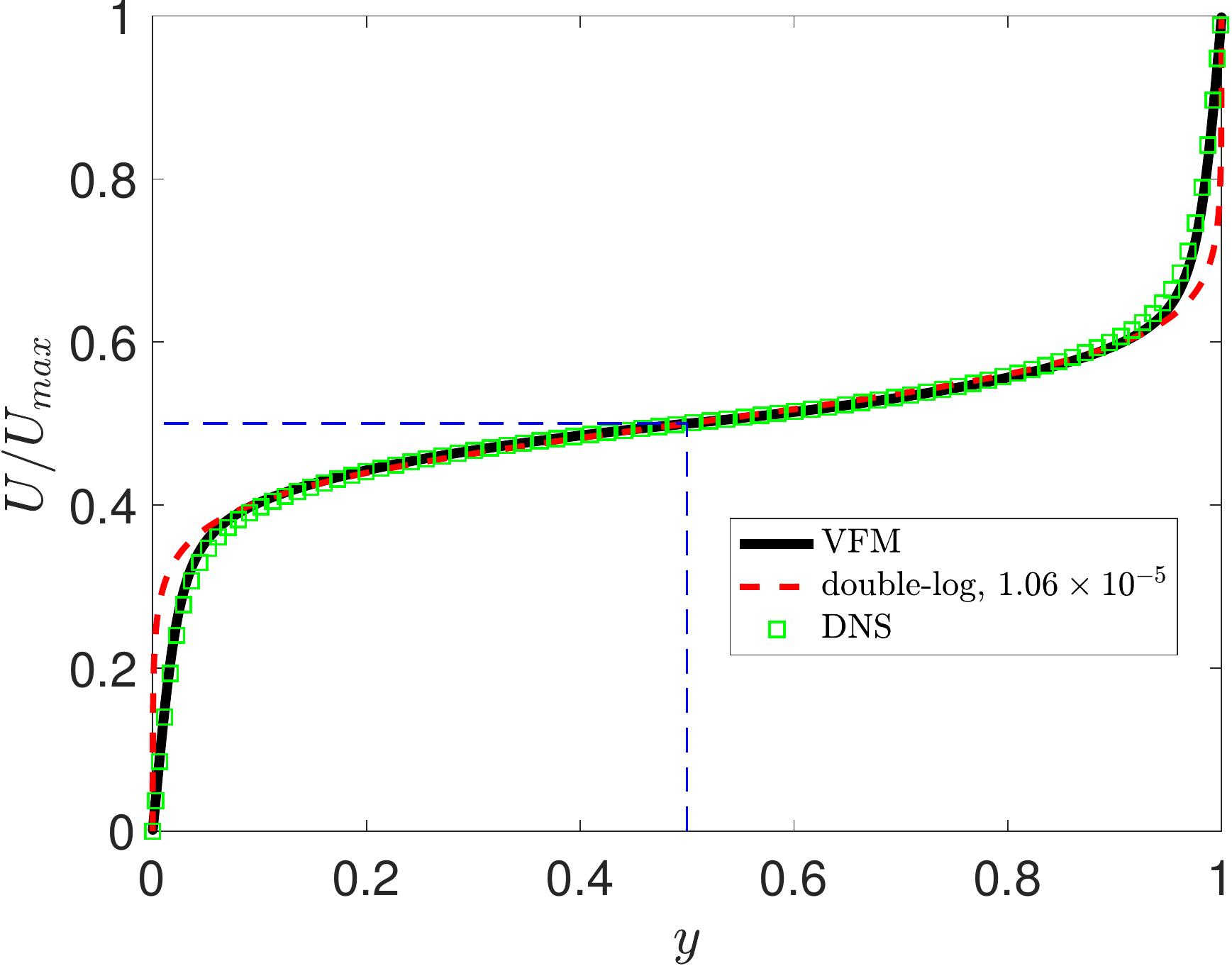}}
\subfloat[]{
\includegraphics[height=2.2in]{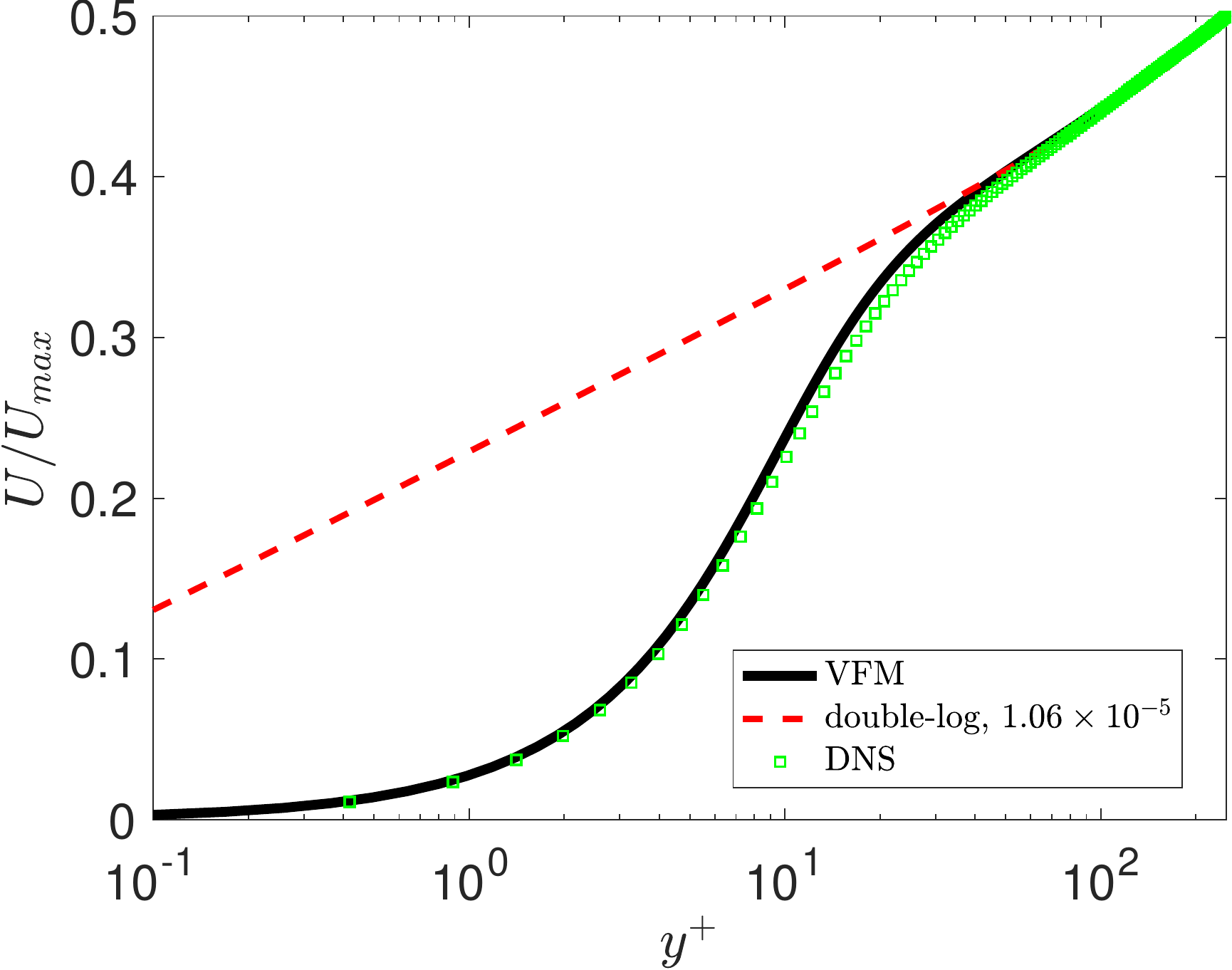}}
\caption{Turbulent Couette flow at $Re_\tau=250$: (a) ``-" VFM predictions, ``- -" best fit of the double-log profile in equation \eqref{lppre} with $d=1.06\times10^{-5}$,  ``$\square$" DNS data from \cite{avsarkisov2014turbulent}; (b) Wall units scaling for the mean velocity profiles.}
\label{profile_y250}
\end{figure}

Here, we consider the predictions from the universal scaling fractional order $\alpha^*(y^+)$, and we also make comparisons against the double-log profile.
The variable fractional order $\alpha^*(y^+)$ is between zero to one in our turbulence model. The fractional equation is not well-posed in the domain
$[0,1]$ so we work in the half plane $y\in[0,0.5]$ (see the dash square in Fig. \ref{profile_y} (a)). We then obtain the results in the other half of
the domain with $U(y)=1-U(1-y), y\in(0.5,1]$.
Fig. \ref{profile_y}  shows the mean velocity profiles predicted using \eqref{lppre} and the mean velocity which is predicted by the variable fractional order $\alpha^*(y^+)$. We can observe that the variable fractional model is in agreement with the experiment data as well as the double-log profile. However, the double-log profile is unable to capture the correct mean velocity near the wall. We also plot the profiles for low Reynolds number $Re_\tau=52$,  where here the numerical data is from reference \cite{liu2003turbulent} in Fig. \ref{profile_Re52}. For the double-log profile we could not find a suitable parameter $d$ to obtain a good fit for the low $Re_\tau=52$. Finally, we show the comparisons between the VFM-predicted mean velocities and DNS data at $Re_\tau=125,180,250$ obtained from reference \cite{avsarkisov2014turbulent}. Figs. \ref{profile_y125}-\ref{profile_y250} show that the fractional predictions are correct almost everywhere, especially near the wall regions for high Reynolds numbers.

\begin{figure}[H]
\centering
\subfloat[]{
\includegraphics[height=2.2in]{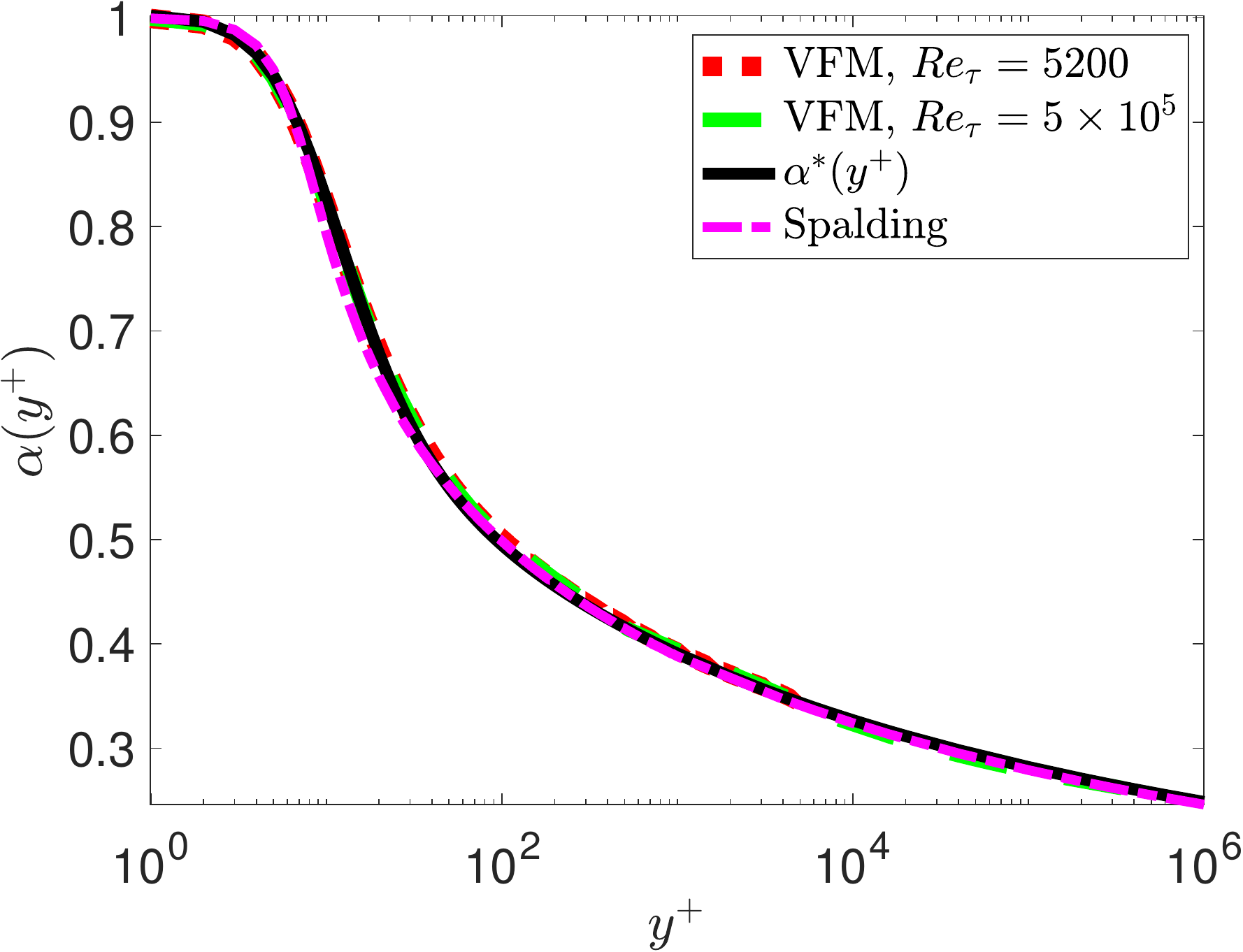}}
\subfloat[]{
\includegraphics[height=2.2in]{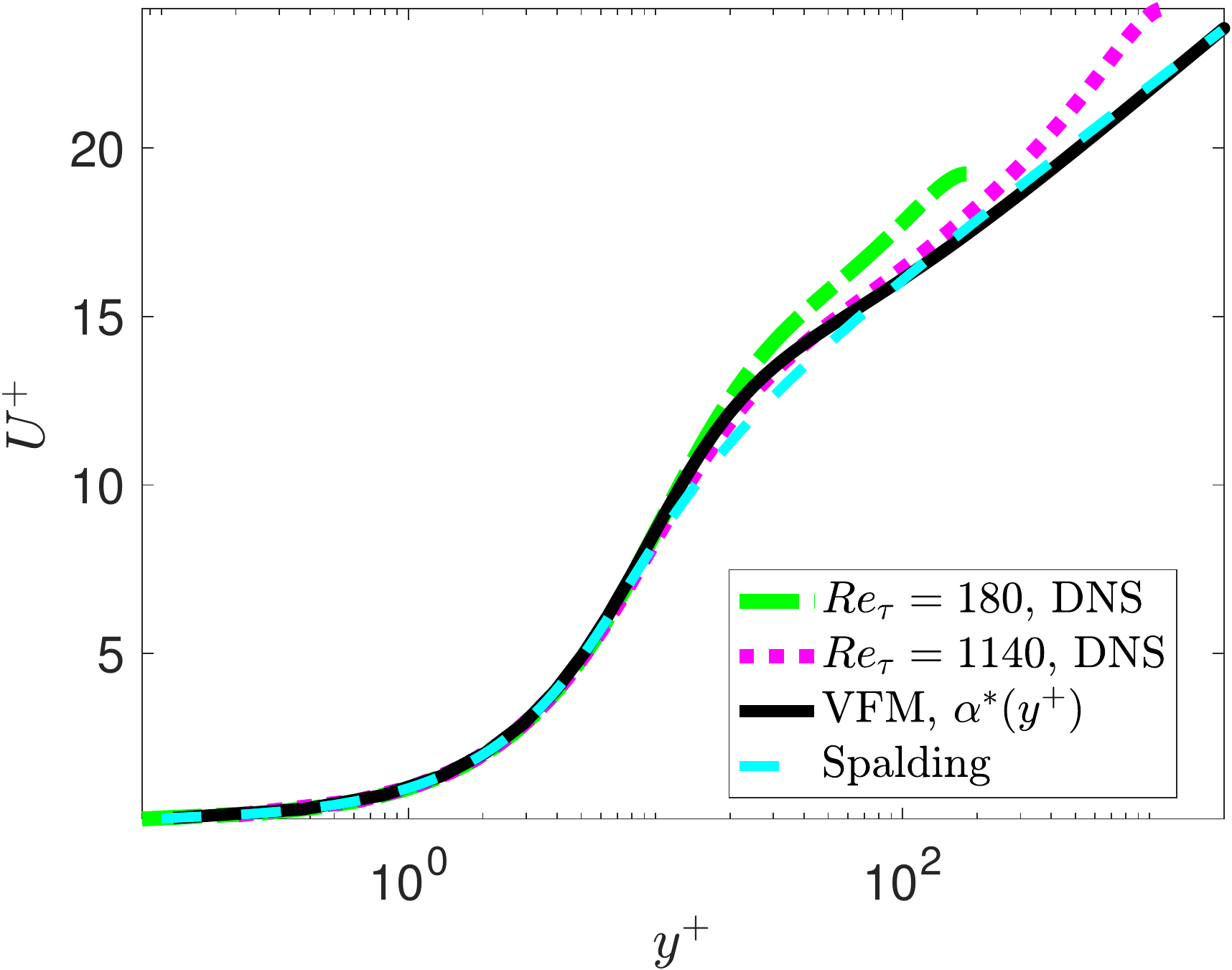}}
\caption{Turbulent pipe flow: (a) ``$\cdot\cdot$" VFM model with the channel flow DNS data at $Re_\tau=5200$, ``- -" VFM model with the M-GPR profile at $Re_\tau=5\times10^5$,  ``-'' the profile of the equation \eqref{polaw} and '-$\cdot$' the corresponding Spalding profile; (b) '-$\cdot$' and '$\cdot\cdot$' plot the DNS data at $Re_\tau=180$ and $Re_\tau=1140$, '-' the VFM model at $Re_\tau=2000$ and the corresponding Spalding profile.}
\label{Re5E5_alpha_py}
\end{figure}

\vspace{0.1in}
{\bf Turbulent Pipe flow --} We first solve the optimization problem using the mean velocity profile obtained from the multifidelity Gaussian process regression (M-GPR) as explained in the main text for superpipe flow at high Reynolds number. Fig. \ref{Re5E5_alpha_py}(a) shows that the variable fractional order we obtain for this problem is identical to the function defined by equation \eqref{polaw}. Fig. \ref{Re5E5_alpha_py}(b) plots the mean velocity profiles from the DNS data  base \cite{wu2008direct} at low Reynolds numbers, the corresponding VFM predictions,  and the Spalding profile.

\begin{figure}[http]
\centering
\subfloat[]{
\includegraphics[height=2.2in]{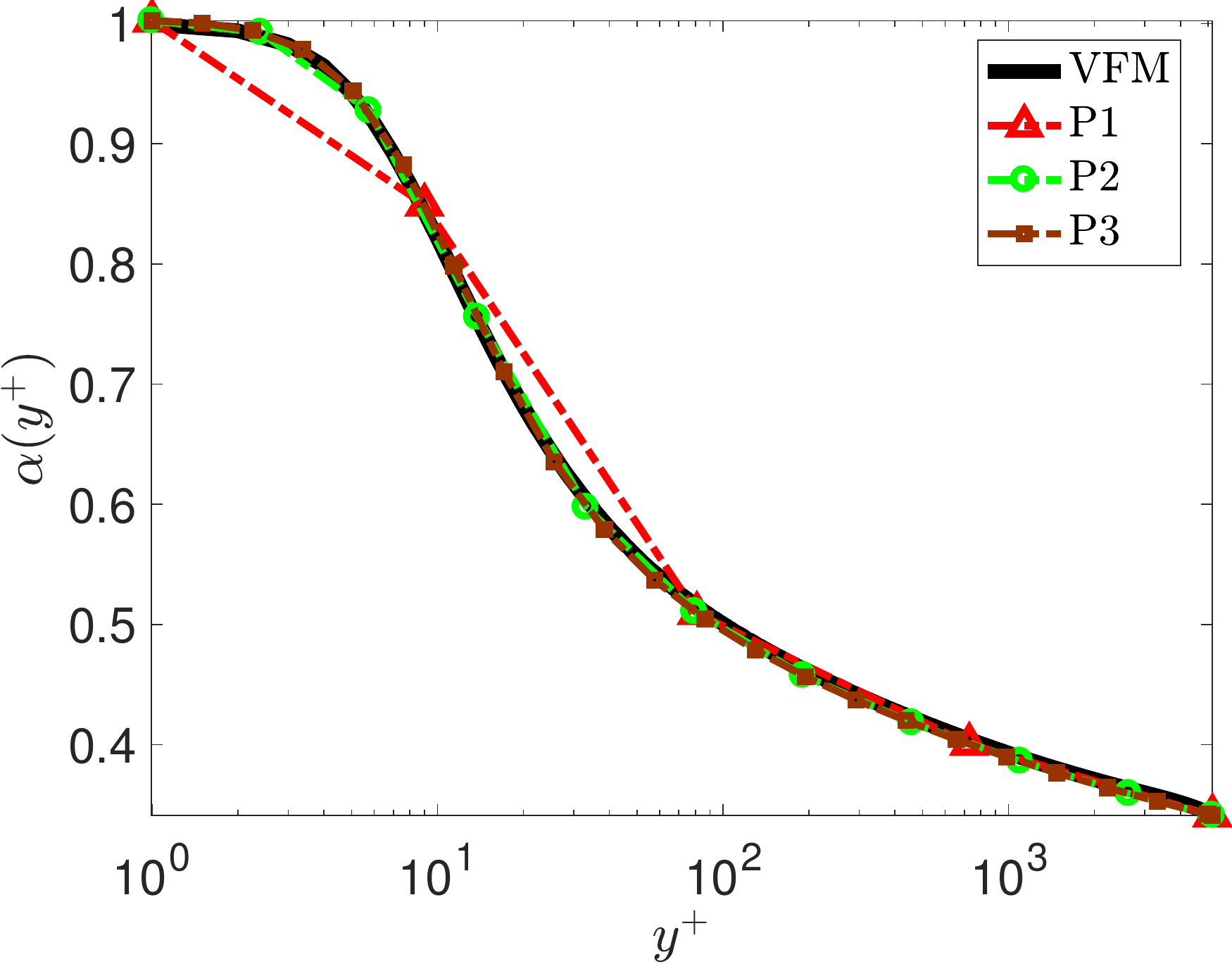}}
\subfloat[]{
\includegraphics[height=2.2in]{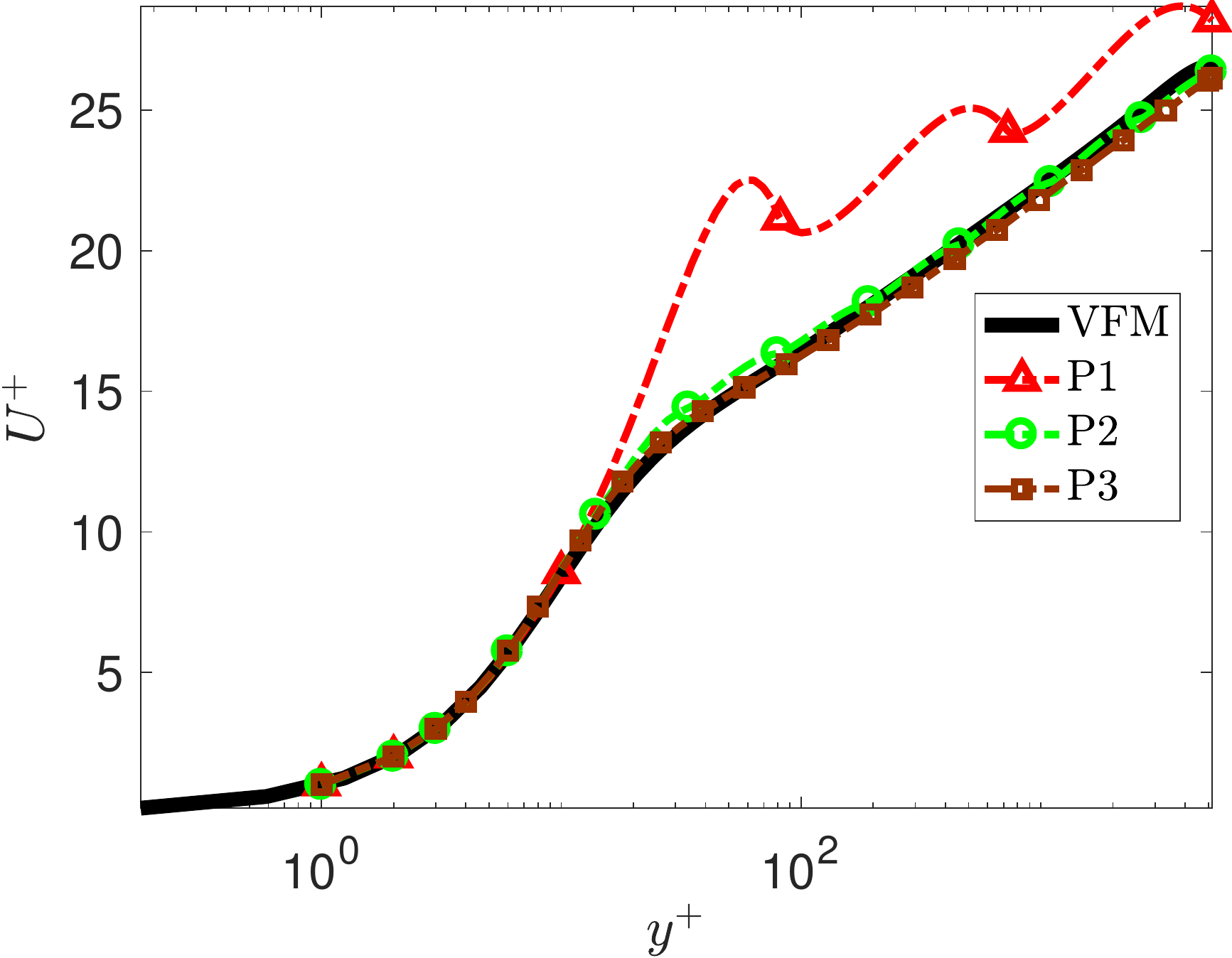}}\\
\caption{Effect of rough approximation of the fractional order: Numerical results of the piecewise linear approximation at $Re_\tau=5200$: (a) $\alpha^*(y+)$ profiles, $P1$ represents 6 points linear interpolation, $P2$ represents 12 points interpolation, and $P3$ represents the numerical results with 24 interpolation points; (b) predicted mean velocity profiles corresponding to the piecewise linear variable order.}
\label{piecelinear}
\end{figure}

\vspace{0.1in}
{\bf Rough approximation of the fractional variable order --} As a last test, we pose the question of how accurately we need to estimate the variable fractional order $\alpha(y)$ in order to obtain accurate results for the mean velocity profile $U(y)$. To this end, we approximate 
the variable fractional order by a piecewise linear function, requiring the solution of an optimization problem on a very sparse grid.  Fig. \ref{piecelinear}  shows the convergence 
of the rough approximations to the very fine fit for turbulent channel flow DNS data at $Re_\tau=5200$. We see that even with 12 interpolation points reasonable accuracy for the mean velocity profile is obtained.  What we envision here is that in multi-dimensional and complex-geometry flows one can sample scattered values of the mean velocity and generate an approximate surrogate surface $\alpha(x,y,z)$ that can be used for solving the variable order fractional Laplacian in three dimensions for data-driven closures and data-driven turbulence simulations. 

\end{document}